\documentclass[structabstract]{aa}
\usepackage{txfonts}
\usepackage {graphicx}
\usepackage{subfig}
\usepackage{color}
\usepackage{amssymb}
\usepackage{ulem}
\usepackage{rotating}
\usepackage{lscape}

\usepackage[]{natbib} 
\usepackage[]{longtable}
\usepackage[]{url}
\bibpunct{(}{)}{;}{a}{}{,}  

\begin{document}


\title{Heavy elements in old very metal-rich stars  
 \thanks{Observations collected at the European  Southern  Observatory, La 
Silla, Chile} }

\titlerunning{Heavy elements in old very metal-rich stars}

\author{
M. Trevisan\inst{1}
\and
B. Barbuy\inst{2}
}
\offprints{M. Trevisan}
\institute{
Instituto Nacional de Pesquisas Espaciais/MCT, S\~ao Jos\'e dos Campos, SP 12227-010, Brazil\\
 e-mail: marina.trevisan@inpe.br
\and
 Universidade de S\~ao Paulo, Rua do Mat\~ao 1226, S\~ao Paulo 05508-900,
 Brazil\\
  e-mail: b.barbuy@iag.usp.br
}

\date{Received; accepted}

 
\abstract
{We studied a sample of high proper motion, old and metal-rich dwarf stars, 
selected from the NLTT catalogue.
The low pericentric distances and eccentric orbits  of these solar neighbourhood stars indicate that
they  might have originated in the inner parts of the Galaxy.}
{Chemical tagging can probe the formation history of stellar populations. To identify the origin of
a sample of 71 very metal-rich dwarf stars,  we derive the abundances of 
the neutron-capture elements Y, Ba, La, and Eu. }
{The spectroscopic analysis is  based on optical high-resolution \'echelle spectra obtained with the  FEROS spectrograph 
at the ESO 1.52-m Telescope at La Silla, Chile. The abundances of Y, Ba, La, and Eu were derived through LTE analysis, 
employing the MARCS model atmospheres.}
{The abundances of Y, Ba, La, and Eu vs. Fe and Mg as the reference elements indicate similarities between our sample of old metal-rich dwarf stars and the thin disk. On the other hand, the abundance ratios using O as the reference element, as well as their kinematics, suggest that our sample is clearly distinct from the thin-disk stars. They could be old inner thin-disk stars, as suggested previously,
or bulge stars. In either cases they would have migrated 
from the inner parts of the Galaxy to the solar neighbourhood.  } 
{}
\maketitle


\section{Introduction}

 The heavy elements Y, Ba, La, and Eu are among the most well-studied 
neutron-capture elements because they show a number of measurable lines in stellar spectra. Chemical abundances in 
the solar system indicate that, while Y, Ba, and La, are 
dominantly formed through slow neutron-capture process,
Eu is dominantly a rapid neutron-capture element. The $s$- and $r$-process contribution to the solar system abundances
is given  for example in studies by \citet{Seeger.etal:1965}, \citet{Arlandini.etal:1999}, 
\citet{Simmerer.etal:2004}, and \citet{Sneden.etal:2008}. At early times in the Galaxy, heavy elements are expected to
be due to an $r$-process contribution, since there was no time for the main $s$-process to operate,
as first suggested by \citet{Truran:1981}. 
Recently, it was suggested by \citet{Cescutti.Chiappini:2014} 
that $s$-process elements could have been produced 
by massive spinstars at early times in the halo. \citet{Chiappini.etal:2011} and \citet{Barbuy.etal:2014} also suggested to use abundances ratios
of neutron-capture elements in old bulge stars
to determine whether they are products of $r$- or $s$-process.

The sites of $r$-process element production are still not known
\citep[e.g.][]{Schatz.etal:2001b, Wanajo.Ishimaru:2006, Kratz.etal:2007, Thielemann.etal:2011}.
Nucleosynthesis in neutron star mergers for the production of $r$-elements
was favoured in recent works by \citet{Rosswog.etal:2014} and \citet{Wanajo.etal:2014}. They
 confirmed that the so-called strong $r$-process
yields elements with atomic number A~$>$~130 (second 
peak of the heavy elements and above), whereas 
 the so-called weak $r$-process could take place
in neutrino-driven winds of the same neutron star mergers
and produce elements with
50~$<$~A~$<$~130, which includes the first peak of the heavy elements.
Abundance ratios of first- and second-peak elements and $s$-dominant elements such as Ba
over $r$-dominant elements such as Eu can contribute to distinguishing different nucleosynthesis
process contributions \citep[e.g.][]{SiqueiraMello.etal:2014, SiqueiraMello.Barbuy:2014}.

In the seminal work by \citet{Kappeler.etal:1989} and \citet{Kappeler.etal:2011}
 three different channels for the
production $s$-process elements are described:
the weak $s$-process in He-burning and C-burning shells of massive stars, and the main and the strong $s$-process 
taking place in asymptotic giant branch (AGB) stars.
These processes were inspected in detail by 
\citet{Gallino.etal:1998} and \citet{Travaglio.etal:2004}, 
 among others.
 
Heavy-element abundances have been examined more frequently in metal-poor than in metal-rich stars. 
Previous heavy-element abundance derivations in metal-rich stars were carried out for thin-disk 
\citep{Edvardsson.etal:1993, Reddy.etal:2003, AllendePrieto.etal:2004}, and thick-disk stars 
\citep{Reddy.etal:2006, Bensby.etal:2005}.
Recently, \citet{Mishenina.etal:2013} derived heavy-element abundances of thin-, thick-disk and Hercules-stream 
stars, and abundances of bulge-field stars were determined by \citet{Bensby.etal:2013} and 
\citet{Johnson.etal:2012, Johnson.etal:2013}.  In addition, heavy-element abundances were derived for 
open-cluster stars \citep[e.g.][]{Jacobson.Friel:2013} and for two bulge barium stars by \citet{Lebzelter.etal:2013}.

We here derive Y, Ba, La, and Eu abundances in 71
 metal-rich stars in the solar neighbourhood. The sample stars are high proper motion, old stars, 
 with low maximum heights above the Galactic plane, 
and their kinematics is indicative
of stars wandering from inside the Galactic bar \citep{Raboud.etal:1998}. 
\citet[][hereafter Paper~I]{Trevisan.etal:2011} presented the detailed analysis of the sample and
derived their $\alpha$-element abundances.

Based on their kinematical properties, in Paper~I we identified the present sample as partly belonging 
to the thin disk (11 stars),
a majority of them 
 to the thick disk (42 stars), and 17 stars to an intermediate population.
The thick-disk and intermediate populations might be similar to the
thick-disk-kinematics-thin-disk-abundances (designated TKTA) stars \citep{Reddy.etal:2006}, which show
 thick-disk kinematics combined with thin-disk abundances.

To probe the formation history of this sample, 
in Paper~I we
derived their metallicities, $\alpha$-element abundances
 (O, Mg, Si, Ca, and Ti), ages, and Galactic orbits.
We showed that models of radial mixing and dynamical effects of the bar and bar/spiral arms might explain the presence of these
old metal-rich dwarf stars in the solar neighbourhood.
In Paper~I we concluded that our sample stars might
 be identified as old thin-disk stars \citep{Haywood:2008}.
To better understand the properties of these stars,
we here study the abundances of the neutron-capture elements
Y, Ba, La, and Eu. 

The paper is organized as follows: in Sect.~\ref{Sec_sample}, we  briefly describe the sample and the derivation of stellar parameters that were presented
in Paper~I. 
The Y, Ba, La, and Eu abundance derivation is described in Sect.~\ref{Sec_abons}, and the results are presented and discussed
in Sects. \ref{Sec_results} and \ref{Sec_discussion}. Finally, the results are summarized in Sect.~\ref{Sec_summary}.


\section{Sample}
\label{Sec_sample}

The sample star selection, 
their kinematics, the stellar parameter determination, and the abundance analysis of $\alpha$-elements are described in detail in Paper~I.
Below we summarize the information about the sample stars.

The sample comprises high proper motion and metal-rich dwarf stars from the NLTT catalogue
\citep{Luyten:1980}, as described in \citet{Grenon:1999} and Paper~I. The observations
were carried out in 2000-2002, within an ESO-ON-IAG agreement.
The spectra were obtained using the Fiber-Fed Extended Range Optical Spectrograph (FEROS) \citep{Kaufer.etal:2000}
 at the 1.52~m telescope at ESO, La Silla.
 The total wavelength coverage is 3560-9200 {\rm \AA} with a resolving power
 (R=$\lambda/\Delta\lambda$) = 48~000, and with mean signal-to-noise ratios of $\sim 100$.

\subsection{Galactic orbits and membership assignment}
\label{Sec_orbits}

\citet{Grenon:1999} derived U, V, and W space velocities for the sample stars, which were used 
to calculate the Galactic orbits. The GRINTON integrator was used for this purpose \citep{Carraro.etal:2002, Bedin.etal:2006}.
This code integrates
the orbits back in time for several Galactic revolutions and returns the minimum and maximum distances from the Galactic
centre ($R_{\rm min}$ , $R_{\rm max}$), maximum height from the Galactic plane
($Z_{\rm max}$) and the eccentricity $e$ of the orbit. 

Based on Galactic velocities U, V, W and eccentricities $e$, 
 we estimated in Paper~I the probability of each sample star to belong
 to either the thin or the thick disk. 
The procedure adopted
relies on the assumption that the space velocities of each population (thin disk, thick disk and halo) follow a
Gaussian distribution, with given mean values and dispersions $\sigma_{\rm U}$, $\sigma_{\rm V}$, $\sigma_{\rm W}$.
We used the velocity ellipsoids by \citet{Soubiran.etal:2003} and found that 42 stars in
the sample can be assigned to the thick disk, 11 are more
likely to be thin-disk stars, and the other 17 stars 
are intermediate between thin- and thick-disk components.

\subsection{Stellar parameters}

The effective temperatures $T_{\rm eff}$ were calculated from the $V-K_S$ colour using the \citet{Casagrande.etal:2010} colour-temperature relations. The surface gravities log $g$ were derived using the HIPPARCOS parallaxes, and the stellar ages and masses from the Yonsei-Yale evolutionary tracks \citep[$Y^2$][]{Demarque.etal:2004}. 
Then, fixing $T_{\rm eff}$ and $\log g$, the iron abundances were derived from \ion{Fe}{I} and \ion{Fe}{II} lines through a local thermodynamic equilibrium (LTE) analysis, employing  plane-parallel MARCS model atmospheres \citep{Gustafsson.etal:2008}  with scaled-solar composition. The microturbulence velocity, $v_{\rm t}$, was obtained by imposing constant iron abundance as a function of equivalent width. 
For our cooler stars ($T_{\rm eff} < 5000$~K), a positive trend of [Fe/H] vs. equivalent width was found even when $v_{\rm t} \sim 0$~km~s$^{-1}$. For this reason, 
we considered that $v_{\rm t}$ can be defined as a function of temperature and gravity, 
and using stars with $T_{\rm eff} > 5200$~K, we defined $v_{\rm t} = f(T_{\rm eff}, \log g)$ and then extrapolated this function to
$T_{\rm eff} < 5200$~K. We adopted $0.3$~km~s$^{-1}$ from the extrapolation of our fit. The metallicity was then replaced by the new 
iron abundance, and the procedure was repeated until there were no significant changes in ($T_{\rm eff}$, $\log g$, 
[Fe/H]\footnote{We use the standard notation, i.e., for elements A and B, ${\rm [A / B]} = \log(N_{\rm A} / N_{\rm B})_{\rm star} - \log (N_{\rm A} / N_{\rm B})_{\odot}$
and $\epsilon({\rm A}) = \log (N_{\rm A} / N_{\rm H}) + 12$.}).

The stellar parameters were tested against excitation and ionization equilibria, and $T_{\rm eff}$ and $\log g$ were further adjusted when necessary.  Only two stars in the sample (HD~94374 and HD~182572) required adjustments to be in satisfactory spectroscopic equilibrium. 

The final adopted temperatures, gravities, and metallicities were compared with data available in the literature. For comparison purposes, the parameters ($T_{\rm eff}$, $\log g$, [Fe/H]) were retrieved from the PASTEL catalogue \citep{Soubiran.etal:2010}, which compiles stellar atmospheric parameters obtained from the analysis of high-resolution, high signal-to-noise spectra. The parameters of 38 of our sample stars are available in this catalogue.  Differences between temperatures considered here and those from the PASTEL catalogue do not exceed $2$\%, except for two stars (HD~31827 and HD~35854). Gravities and metallicities also agree well, with differences within $\sim 0.2$~dex.

\begin{figure*}[!ht]
\centering
\begin{tabular}{c}
\vspace{-2.0cm}
 \resizebox{0.95\hsize}{!}{\includegraphics[angle=-90]{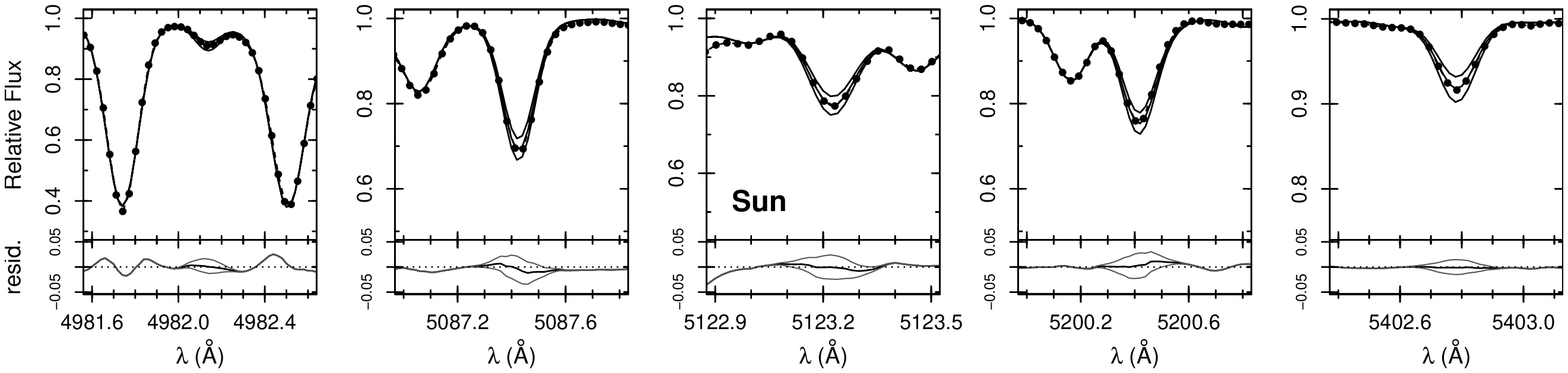}} \\
 \vspace{-1.4cm}
 \resizebox{0.95\hsize}{!}{\includegraphics[angle=-90]{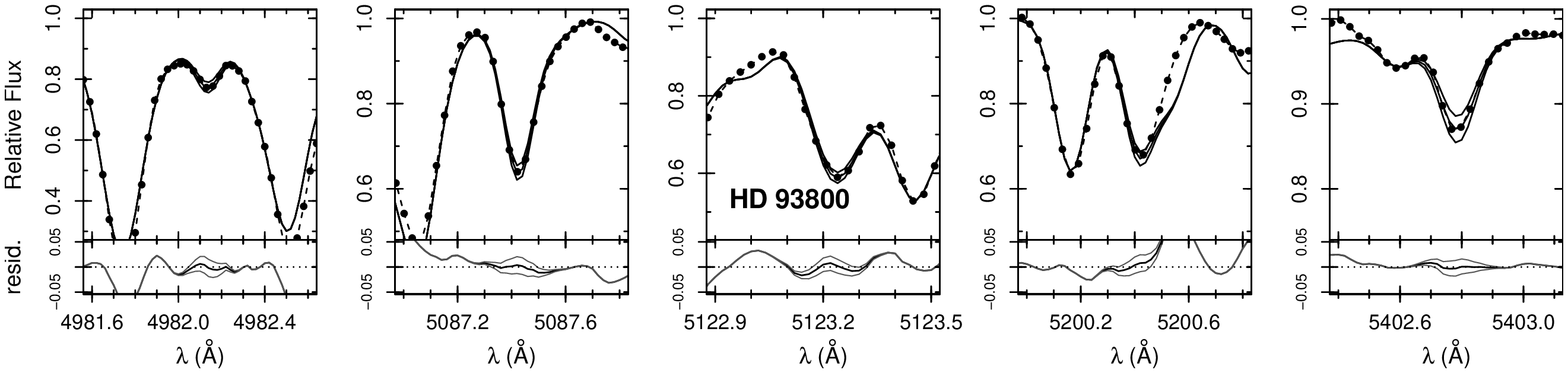}} \\

\end{tabular}
\caption { Synthetic and observed profiles of \ion{Y}{II} lines. {\it Upper panels:} Solar spectrum in the region of \ion{Y}{II} lines. Solid lines indicate the synthetic spectra for $\epsilon({\rm Y})$ $ = 2.14,\ 2.24,\ 2.34$; the dots indicate the observed spectrum. {\it Lower panels:} Spectral synthesis 
of \ion{Y}{II} lines for HD~93800. }
\label{Fig_solar_YLines}
\end{figure*}

\section{Abundance derivation}
\label{Sec_abons}

We derived abundances of yttrium, barium, lanthanum, and europium for the sample stars. We performed a spectral synthesis, and the abundances were obtained by minimising the $\chi^2$ between the observed and synthetic spectra. The synthetic spectra were obtained using the PFANT code described in \citet{Cayrel.etal:1991}, \citet{Barbuy.etal:2003}, and \citet{Coelho.etal:2005}. The MARCS model atmospheres \citep{Gustafsson.etal:2008} were employed. 

The solar abundances from the literature and adopted values are reported in Table \ref{sol}. 
 The most updated values computed with 3D model atmospheres by \citet{Asplund.etal:2009} and \citet{Grevesse.etal:2014} recommended values
 were not adopted here because we used 1D model atmospheres to derive the abundances of the sample stars. 

All lines were checked against the solar spectrum obtained with the same instrumentation as the program stars, 
and gf-values were adjusted when necessary. 
For lines of the elements \ion{Ba}{II}, \ion{La}{II}, and \ion{Eu}{II}, the hyperfine structure (HFS) was taken into account, 
 as described in the following sections.

\begin{table}
\caption{Solar {\it r}- and {\it s}-process fractions \citep{Simmerer.etal:2004} and solar abundances 
from 1: \citet{Grevesse.Sauval:1998}; 2: \citet{Asplund.etal:2009}; 3: \citet{Lodders.etal:2009};
4: \citet{Grevesse.etal:2014}. Adopted abundances are given in Col. 5. }             
\label{sol}      
\centering                          
\begin{tabular}{c c c c c c c c c}        
\hline\hline                 
\hbox{El.} & \hbox{Z} & \multicolumn{2}{c}{Fraction} & \multicolumn{5}{c}{\hbox{$\epsilon$(X)$_{\odot}$}}  \\
 \cline{5-9}   \\
 &  & r & s & (1) & (2)  & (3) & (4) & (5)  \\
\hline                        
\hbox{Fe} & 26 &  ---   &  --- &  7.50  &  7.50  &  7.45  & 7.45 & 7.50  \\
\hbox{Y}  & 39 & 0.281 & 0.719 &  2.24  &  2.21  &  2.21  & 2.21 & 2.24 \\
\hbox{Ba} & 56 & 0.147 & 0.853 &  2.13  &  2.18  &  2.18  & 2.25 & 2.13 \\
\hbox{La} & 57 & 0.246 & 0.754 &  1.17  &  1.10  &  1.14  & 1.11 & 1.22 \\
\hbox{Eu} & 63 & 0.973 & 0.027 &  0.51  &  0.52  &  0.52  & 0.52 & 0.51 \\

\hline                                
\end{tabular}

\end{table}

\subsection*{Yttrium}
\label{Sec_Y}
\begin{figure}[!ht]
\centering
\begin{tabular}{c}
 \resizebox{0.95\hsize}{!}{\includegraphics[angle=0]{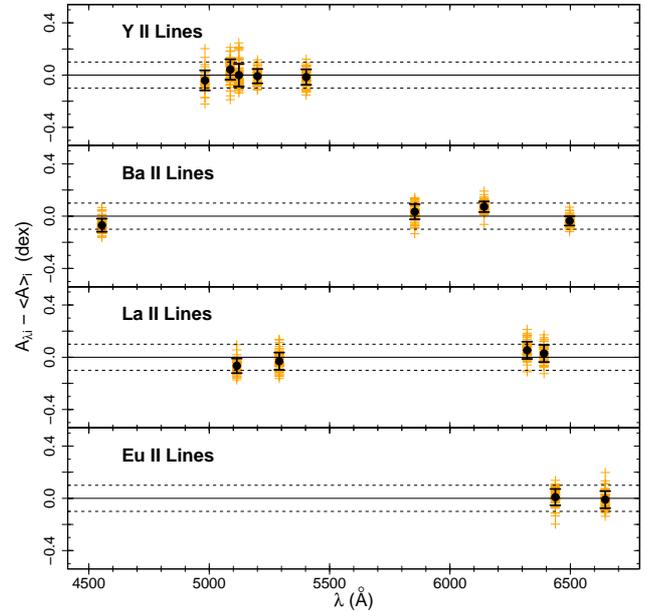}} \\

\end{tabular}
\caption {Selected Y, Ba, La, and Eu lines. The abundance of each star, $\left\langle A \right\rangle_i$, is the average
of abundances derived from individual lines, $A_{\lambda i}$. Therefore, $A_{\lambda i} - \left\langle A \right\rangle_i$ 
indicates the differences between the mean abundance value and those derived from individual lines. The mean deviation for each line
is represented by black circles. Dashed lines indicate $\pm 0.1$~dex.}
\label{Fig_YBaLaEuLines}
\end{figure}

The \ion{Y}{II} lines were selected from \citet{Hannaford.etal:1982}, \citet{Spite.etal:1987}, \citet{Hill.etal:2002},
and \citet{Grevesse.etal:2014}, and they are listed in Table \ref{Tab_linelist}.
 The $\log gf$ values were adjusted to fit the solar spectrum. 
Figure \ref{Fig_solar_YLines} presents the solar observed and synthetic spectra in the region of \ion{Y}{II} lines. The spectrum synthesis fitting of \ion{Y}{II} lines to the observed profiles for one of the sample stars, HD~93800, is also presented in Fig.~\ref{Fig_solar_YLines}.
The yttrium line list comprises lines that systematically give reliable abundances, 
as shown in Fig.~\ref{Fig_YBaLaEuLines}. Given that $\left\langle A \right\rangle_i$ is the Yttrium abundance of each star and  $A_{\lambda i}$ is the abundance derived from an individual line, the quantity 
$A_{\lambda i} - \left\langle A \right\rangle_i$ indicates the departure of the abundance derived from the line $i$.  
This approach allows us to detect lines that give abundances systematically higher or lower than the average abundance of the star and lines that seem to lead to inaccurate abundance values.  The abundances $A_{\lambda i}$ presented in Fig. \ref{Fig_YBaLaEuLines} correspond to the corrected values obtained 
by removing the dependence with  temperature (see Sect. \ref{Sec_trends}). 

\begin{table*}[!ht]
\centering
\footnotesize
 \caption{Line list}
\begin{tabular}{cccccccccccc}
\hline\hline
$\lambda$ (\AA) & Species & $\chi_{exc}$  & $\log gf$ & $\log gf$  & $\log gf$ &  $\log gf$  & $\log gf$ &  $\log gf$  &  $\log gf$ &  $\log gf$&  $\log gf$  \\
                &         &    (eV)       &  (adopted)    &  (HPC+02)  & (SHFS87)  &  (HLG+82)  & (GSAS14)  & (NIST) & (Kurucz)    & (VALD) & (MPK13)      \\ 
\hline

     4982.14 &  Y II &  1.03 & -1.34 &  ---  & -1.26 & -1.29  &   --- & --- & -1.29 &  -1.29 & -1.26 \\
     5087.43 &  Y II &  1.08 & -0.38 & -0.17 & -0.20 & -0.17  & -0.17 & --- & -0.17 &  -0.17 & -0.26 \\
     5123.21 &  Y II &  0.99 & -0.83 & -0.83 &   --- & -0.83  &   --- & --- & -0.83 &  -0.83 &  ---  \\
     5200.41 &  Y II &  0.99 & -0.62 & -0.57 &   --- & -0.57  & -0.57 & --- & -0.57 &  -0.57 & -0.63 \\

\vspace{0.2cm}
     5402.78 &  Y II &  1.84 & -0.58 &   --- & -0.59 &   ---  &   --- & --- & -0.51 &  -0.51 & -0.55   \\
     
     4554.03 & Ba II &  0.00 & +0.30$^a$ & +0.17 &  ---  & --- & +0.17    & +0.14  & +0.17  & +0.17  &  --- \\
     5853.69 & Ba II &  0.60 & -0.90$^a$ & -1.01 & -0.53 & --- & -1.03    & -0.91  & -1.00  & -1.00  &  --- \\
     6141.71 & Ba II &  0.70 & +0.15$^a$ & -0.07 & +0.40 & --- & ---      & -0.03  & -0.08  & -0.08  &  --- \\
\vspace{0.2cm}
     6496.91 & Ba II &  0.60 & -0.10$^a$ &  ---  & +0.18 & --- & -0.41    & -0.41  & -0.38  & -0.38  &  --- \\

     5114.55 & La II &  0.24 & -1.02$^a$ & ---   & ---   & --- &  ---     & ---    & -1.06  & -1.03  & ---   \\
     5290.82 & La II &  0.00 & -1.70$^a$ & ---   & -1.29 & --- &  ---     & ---    & -1.75  & -1.65  & ---   \\
     6320.38 & La II &  0.17 & -1.32$^a$ & -1.52 & -1.43 & --- & ---      & ---    & -1.61  & -1.56  & -1.33 \\  
\vspace{0.2cm}
     6390.48 & La II &  0.32 & -1.31$^a$ & ---   & ---   & --- &  ---     & ---    & -1.45  & -1.41  & ---   \\

     6437.64 & Eu II &  1.32 & +0.26$^a$ & -0.32 & ---   & --- & ---      & ---    & -0.28  & -0.32  & ---   \\
\vspace{0.2cm}
     6645.11 & Eu II &  1.38 & +0.20$^a$ & +0.12 & ---   & --- &  ---     & ---    & +0.20  & +0.12  & ---  \\

\hline
\end{tabular}
\label{Tab_linelist}
\tablefoot{
  HPC+02: \citet{Hill.etal:2002}; SHFS87: \citet{Spite.etal:1987}; HLG+82: \citet{Hannaford.etal:1982}; 
 GSAS14: \citet{Grevesse.etal:2014}; MPK13: \citet{Mishenina.etal:2013}.
 \tablefoottext{a}{$\log gf$ reported corresponds to the sum of individual values.}
}
\end{table*}

\begin{figure*}[!ht]
\centering
\begin{tabular}{c}
\vspace{-2.8cm}
 \resizebox{0.95\hsize}{!}{\includegraphics[angle=-90]{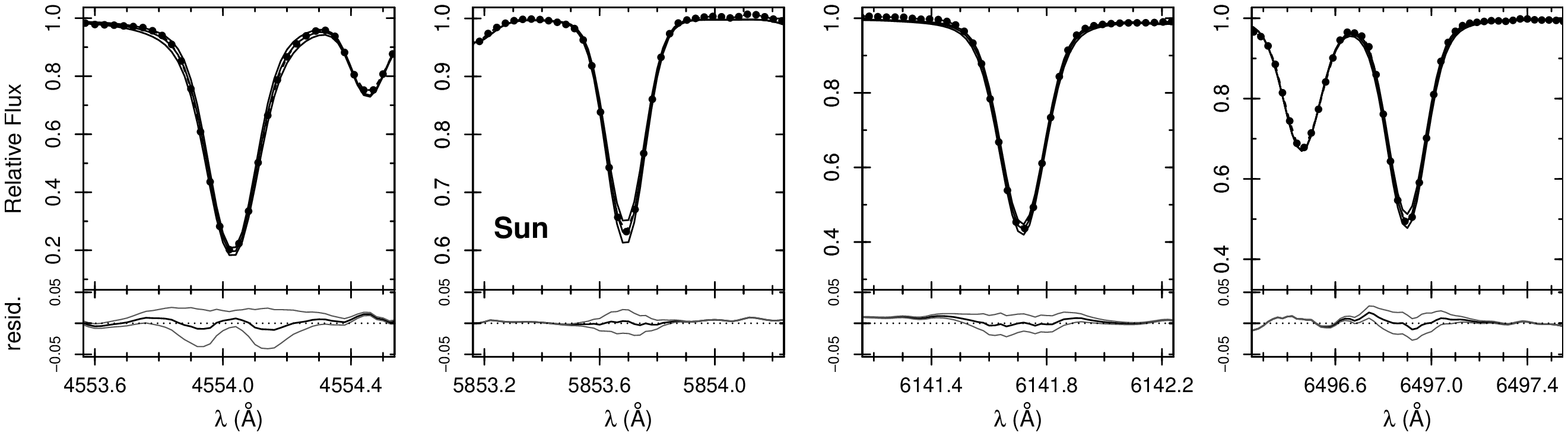}} \\
 \vspace{-2cm}
 \resizebox{0.95\hsize}{!}{\includegraphics[angle=-90]{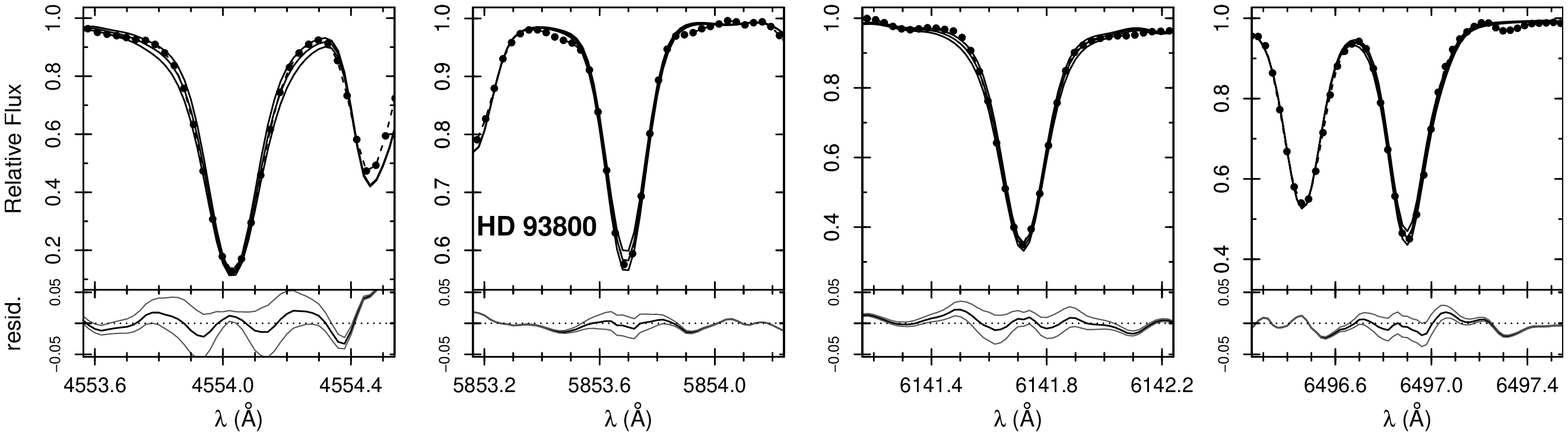}} 
\end{tabular}
\caption {Spectra in the region of \ion{Ba}{II} lines. 
{\it Upper panels:} Solar spectrum. Solid lines indicate the synthetic spectra for 
$\epsilon({\rm Ba})$ $ = 2.03,\ 2.13,\ 2.23$; the observed spectrum is represented by dots. {\it Lower panels:}  Spectrum synthesis fitting
of \ion{Ba}{II} lines for HD~93800. }
\label{Fig_solar_BaLines}
\end{figure*}

\begin{figure*}[!ht]
\centering
\begin{tabular}{c}
\vspace{-2.8cm}
 \resizebox{0.95\hsize}{!}{\includegraphics[angle=-90]{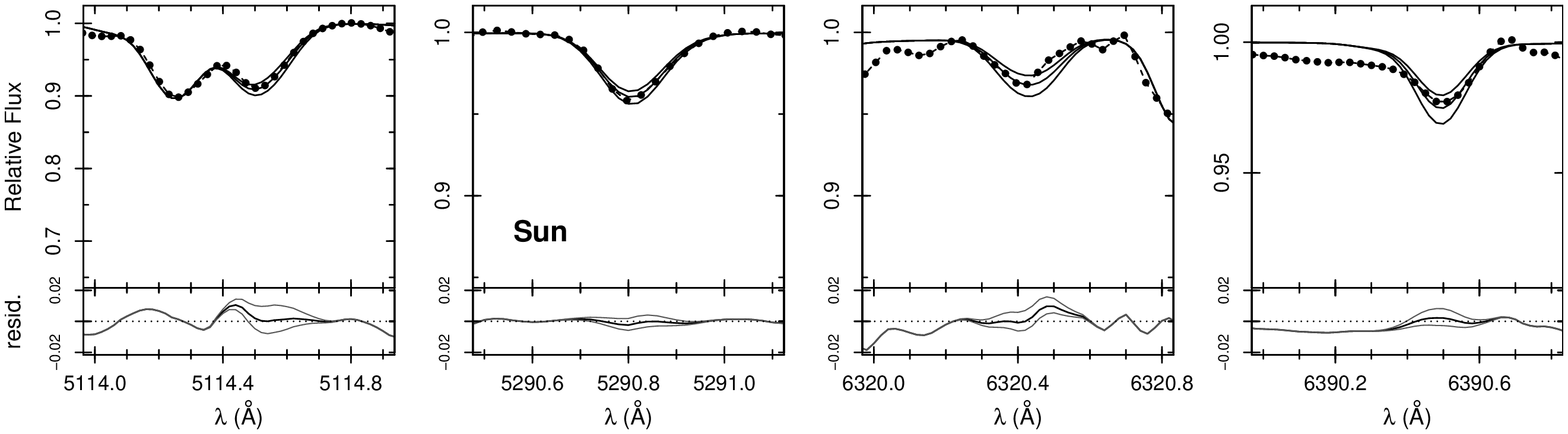}} \\
 \vspace{-2.0cm}
\resizebox{0.95\hsize}{!}{\includegraphics[angle=-90]{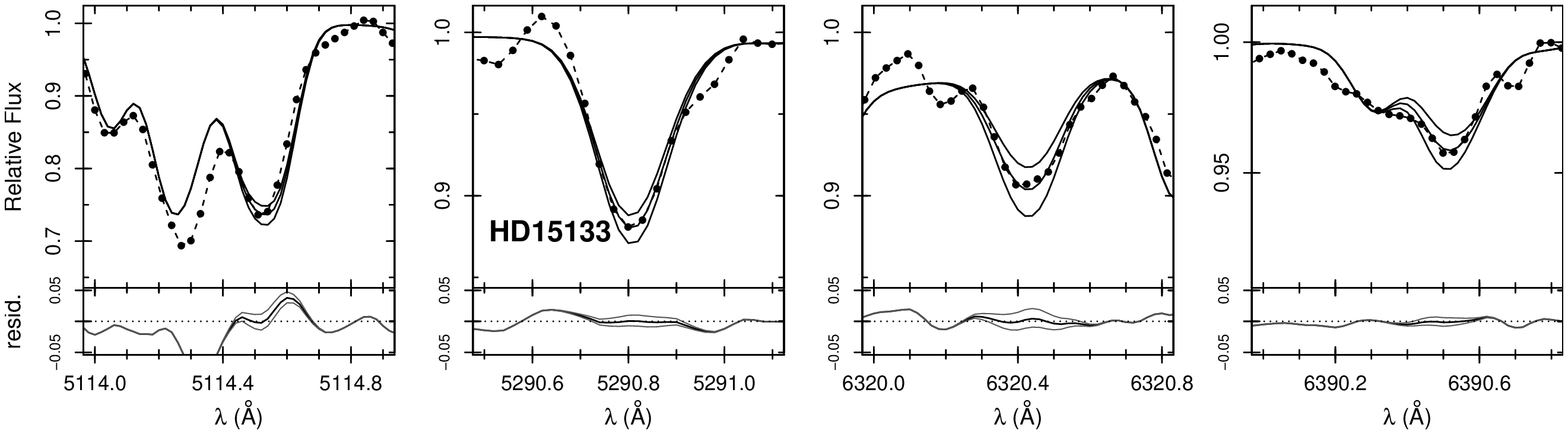}} 

\end{tabular}
\caption { Spectra in the region of \ion{La}{II} lines. {\it Upper panels:}
Observed and synthetic solar spectra.
Solid lines indicate the synthetic spectra for $\epsilon({\rm La})$ $ = 1.12,\ 1.22,\  1.32$; 
the dots indicate the observed spectrum. {\it Lower panels:} Spectrum synthesis fitting
for HD~15133. } 
\label{Fig_solar_LaLines}
\end{figure*}

\subsection*{Barium}

The barium abundances were obtained using  \ion{Ba}{II} lines at  $4554$, $5853$, $6141$, and $6496$~\AA, as listed in Table \ref{Tab_linelist}. 
 Odd-numbered Ba isotopes exhibit hyperfine splitting, which was taken into account by employing a code made available by \citet{McWilliam:1998}, following the calculations described by \citet{Prochaska.McWilliam:2000}, as reported in \citet{Barbuy.etal:2014}.
 We considered that the nuclides $^{134}$Ba, $^{135}$Ba, $^{136}$Ba, $^{137}$Ba, $^{138}$Ba 
contribute 2.42\%, 6.59\%, 7.85\%, 11.23\%, and 71.7\% to the total   abundance, respectively \citep{Lodders.etal:2009}.
Figure \ref{Fig_solar_BaLines} shows the spectra of the Sun and the star HD~93800 in the region of \ion{Ba}{II} lines. 
To fit the solar spectrum, the total $\log gf$ of the Ba lines was increased by $\sim 0.1-0.2$~dex with respect
to values from the literature. The source of this difference is the adopted value 
for the solar microturbulence velocity; we used the same value as in Paper~I ($v_{\rm t, \odot} = 0.90$~km~s$^{-1}$).
We checked the reliability of abundances derived from individual Ba lines using the same procedure as applied for Y lines,
as presented in Fig. \ref{Fig_YBaLaEuLines}. 
The mean deviation of abundances derived from each Ba line do not exceed $0.1$~dex.

\subsection*{Lanthanum}

The La abundances were obtained using the lines listed in Table~\ref{Tab_linelist}. 
The solar and HD~15133 spectra in the region of La lines are shown in Fig.~\ref{Fig_solar_LaLines}.
The HFS of the \ion{La}{II} line at 5114~\AA\ was taken from \citet{Ivans.etal:2006}. 
For lines at 5290~\AA, 6320~\AA\ and 6390~\AA, the HFS was obtained by using the 
code made available by \citet{McWilliam:1998} \citep[see][]{Prochaska.McWilliam:2000}, as reported in \citet{Barbuy.etal:2014}. 
The nuclide $^{139}$La is the dominant isotope, contributing 99.9\% to the total solar system abundance \citep{Lawler.etal:2001a}.

For the \ion{La}{II} line at $6320$~\AA, the effect of a \ion{Ca}{I} line at  $6318.3$~\AA\ that shows autoionization effects and produces a $\sim$5~\AA\ broad line was taken into account \citep[e.g.][see also the derivation of Mg abundances in Paper~I]{Lecureur.etal:2007}. 
The \ion{Ca}{I} autoionization line was treated by increasing its radiative broadening to reflect the much shorter lifetime of the level that is autoionized compared with the radiative lifetime of this level. The radiative broadening had to be increased by 16~000 compared with its standard value ($\propto 1/ \lambda^2$, based on the radiative lifetimes alone) to reproduce the \ion{Ca}{I} dip in the solar spectrum. The Ca abundance for each star was taken into account when computing the \ion{Ca}{I} line in the 6320 \AA\ region.

\subsection*{Europium}

\begin{figure}[!ht]
\centering
\begin{tabular}{c}
\vspace{-2.2cm}
 \resizebox{0.9\hsize}{!}{\includegraphics[angle=-90]{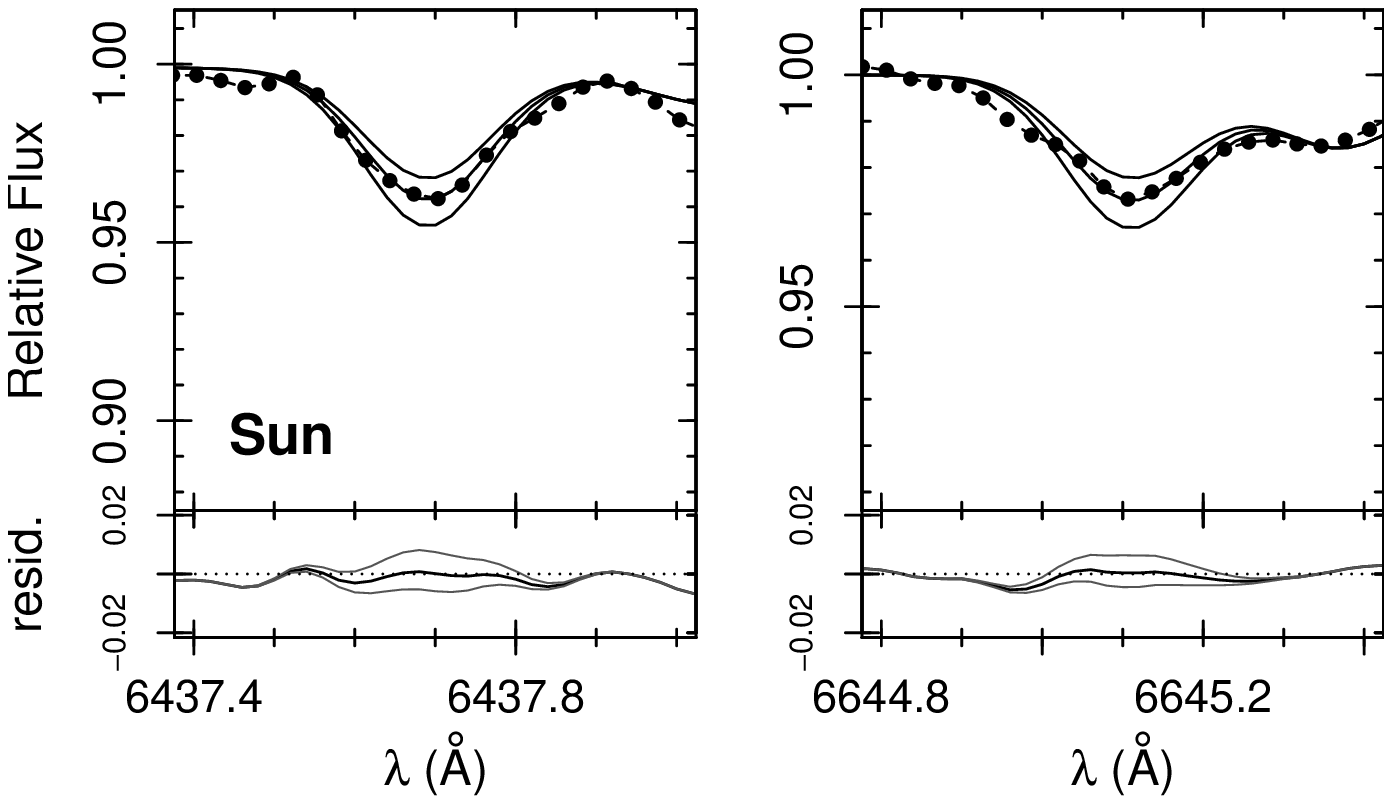}} \\
 \vspace{-1.8cm}
 \resizebox{0.9\hsize}{!}{\includegraphics[angle=-90]{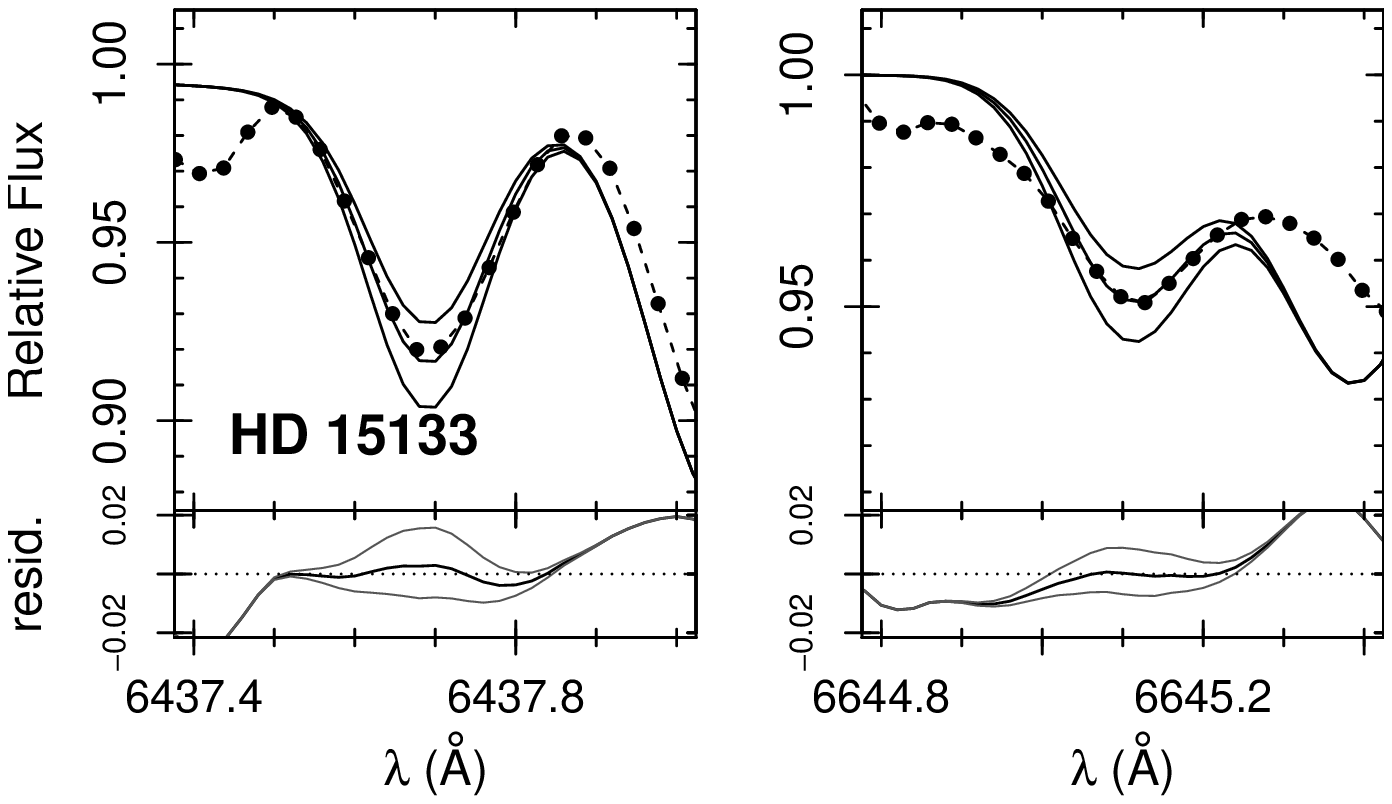}} 

\end{tabular}
\caption {Profiles of the \ion{Eu}{II} line at $6437$~\AA\ and $6645$~\AA. {\it Upper panels:} Solar spectrum. Solid lines indicate the synthetic spectra for $\epsilon({\rm Eu})$ $ = 0.41,\ 0.51,\  0.61$; the dots indicate the observed spectrum. {\it Lower panels:} Spectrum synthesis fitting
of \ion{Eu}{II} lines for HD~15133. } 
\label{Fig_solar_EuLines}
\end{figure}

Europium abundances were derived using \ion{Eu}{II} lines at $6437$~\AA\ and $6645$~\AA. 
 We used the HFS given by \citet{Ivans.etal:2006},  
and isotopic proportions of 47.8\% for  $^{151}$Eu and 52.2\% for $^{153}$Eu \citep{Lawler.etal:2001b}.
 The total $\log gf$ value was adjusted to fit the solar spectrum (Fig.~\ref{Fig_solar_EuLines}) by 
 adopting $\epsilon({\rm Eu})$~=~0.51 \citep{Grevesse.Sauval:1998}. 
The observed and the synthetic spectra of HD~15133 in the region of \ion{Eu}{II} are also presented in Fig.~\ref{Fig_solar_EuLines}.

\subsection{Spurious abundance trends}
\label{Sec_trends}

\begin{figure}
\centering
\begin{tabular}{cc}
 \resizebox{\hsize}{!}{\includegraphics{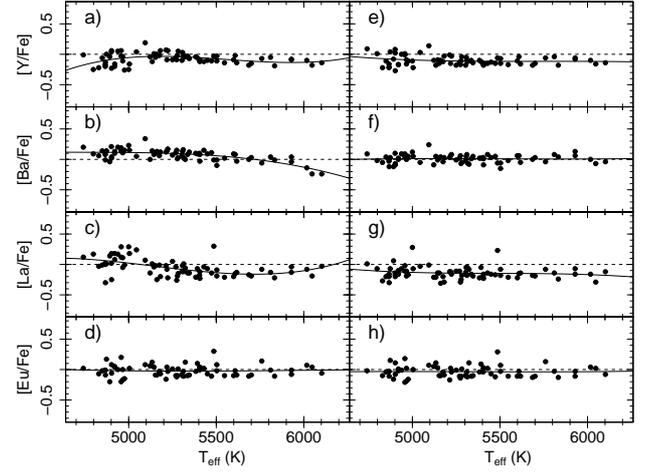}}
 
 \end{tabular}
\caption {Abundances of Y, Ba, La, and Eu {\it vs.} temperature before ({\it panels a, b, c,} and {\it d}) and after ({\it panels e, f, g,} and {\it h}) the correction for the trend with $T_{\rm eff}$. We considered the mean abundance in the range  $5680 < T_{\rm eff} < 5880$ K as the zero point of the correction. The curve in the left panels is shown to illustrate the procedure, which was applied 
for abundances individually derived from each line.}
\label{Fig_teff_abonds}
\end{figure}

We determined the dependence of our final abundances on $T_{\rm eff}$. 
Panels a, b, c, and d  in Fig.~\ref{Fig_teff_abonds} show the abundances {\it vs.} $T_{\rm eff}$, 
and a significant trend is observed for Y, Ba, and La abundances. Following a procedure similar to that adopted in Paper~I, 
we corrected for this trend by fitting a second-order polynomial and by assuming that abundance trend corrections are null at $T_{\rm eff}~=~5777$~K. 
This procedure was applied for abundances individually derived from each line , and the corrected abundances {\it vs.} $T_{\rm eff}$  
are shown in panels e, f, g, and h in Fig.~\ref{Fig_teff_abonds}.

%

\subsection{Errors}

 The errors in abundances result mainly from the errors of the stellar parameters and from the errors in fitting the synthetic spectra.
We estimated the errors in abundances due to uncertainties in the stellar parameters, as presented in Table
\ref{Tab_errors}. The variations $\Delta$[Y, Ba, La, Eu/H] were computed by changing the atmospheric parameters by $\Delta T_{\rm eff} = +100$~K, 
$\Delta \log g = +0.2$~dex and $\Delta v_{\rm t} = +0.2$~km~s$^{-1}$. This procedure was performed for three of the sample stars with different temperatures
($T_{\rm eff} \sim 5900$~K and $\sim 5200$~K) and metallicities ([Fe/H]~$\sim 0.2$ and $\sim 0.5$).

Errors in the final abundance of Y and Eu are $\sim 0.06-0.08$~dex, and errors in [La/H] are $\sim 0.10-0.12$ for the three stars.  
Uncertainty in the Ba abundance reach $0.19$~dex for the star with high temperature, but it is smaller for cool stars ($\sim 0.05-0.11$~dex).

The final abundances of Y, Ba, La and Eu are given in Table \ref{Tab_final_abonds}.
The uncertainties presented in this Table correspond to line-to-line scatter.

\begin{table}[!ht]
\centering
\tiny
 \caption{Abundance variations with stellar parameters.}
\begin{tabular}{lcccc}
\hline\hline
 & \multicolumn{4}{l}{HD 82943 } \\
 & \multicolumn{4}{l}{($T_{\rm eff} = 5929$~K, $\log g = 4.35$, $v_{\rm t} = 1.22$~km~s$^{-1}$, [Fe/H]$= 0.23$)} \\ 

 \cline{2-5}   \\ 
                   &       $\Delta {\rm T}_{\rm eff}$ &  $\Delta \log g$ & $\Delta v_{\rm t}$  & Total \\

                   &                 ($100$~K)        &     ($0.2$~dex)  & ($0.2$~km~s$^{-1}$) &       \\

\hline                                   
                                                                                               
$\Delta$[Y/H]   & $+0.01$ & $+0.07$ & $-0.03$ & $+0.07$ \\                                                                                                                                                                         
$\Delta$[Ba/H]  & $+0.04$ & $+0.04$ & $-0.18$ & $+0.19$ \\                                                                                                             
$\Delta$[La/H]  & $+0.05$ & $+0.09$ & $-0.00$ & $+0.10$ \\
$\Delta$[Eu/H]  & $-0.01$ & $+0.07$ & $-0.01$ & $+0.07$ \\

\hline
          & \multicolumn{4}{l}{HD 87007 } \\ 
          & \multicolumn{4}{l}{($T_{\rm eff} = 5282$~K, $\log g = 4.54$, $v_{\rm t} = 0.61$~km~s$^{-1}$, [Fe/H]$= 0.29$)} \\
  \cline{2-5}   \\ 
                   &       $\Delta {\rm T}_{\rm eff}$ &  $\Delta \log g$ & $\Delta v_{\rm t}$  & Total \\

\hline
$\Delta$[Y/H]   & $+0.01$ & $+0.07$ & $-0.04$ & $+0.08$ \\          
$\Delta$[Ba/H]  & $+0.03$ & $+0.02$ & $-0.10$ & $+0.11$ \\ 
$\Delta$[La/H]  & $+0.04$ & $+0.11$ & $+0.01$ & $+0.12$ \\ 
$\Delta$[Eu/H]  & $-0.01$ & $+0.06$ & $+0.01$ & $+0.06$ \\   
\hline
          & \multicolumn{4}{l}{HD 93800 } \\
          & \multicolumn{4}{l}{($T_{\rm eff} = 5181$~K, $\log g = 4.44$, $v_{\rm t} = 0.30$~km~s$^{-1}$, [Fe/H]$= 0.49$)} \\
\cline{2-5}   \\ 
                   &       $\Delta {\rm T}_{\rm eff}$ &  $\Delta \log g$ & $\Delta v_{\rm t}$  & Total \\
\hline
               
$\Delta$[Y/H]   & $+0.03$ & $+0.07$ & $-0.04$ & $+0.08$ \\
$\Delta$[Ba/H]  & $+0.02$ & $+0.00$ & $-0.05$ & $+0.05$ \\
$\Delta$[La/H]  & $+0.04$ & $+0.10$ & $+0.01$ & $+0.11$ \\ 
$\Delta$[Eu/H]  & $-0.01$ & $+0.07$ & $-0.01$ & $+0.07$ \\

\hline
\end{tabular}
\label{Tab_errors}

\end{table}

\section{Results}
\label{Sec_results}

\begin{figure*}[!ht]
\centering
\begin{tabular}{c}
 \resizebox{\hsize}{!}{\includegraphics[angle=-90]{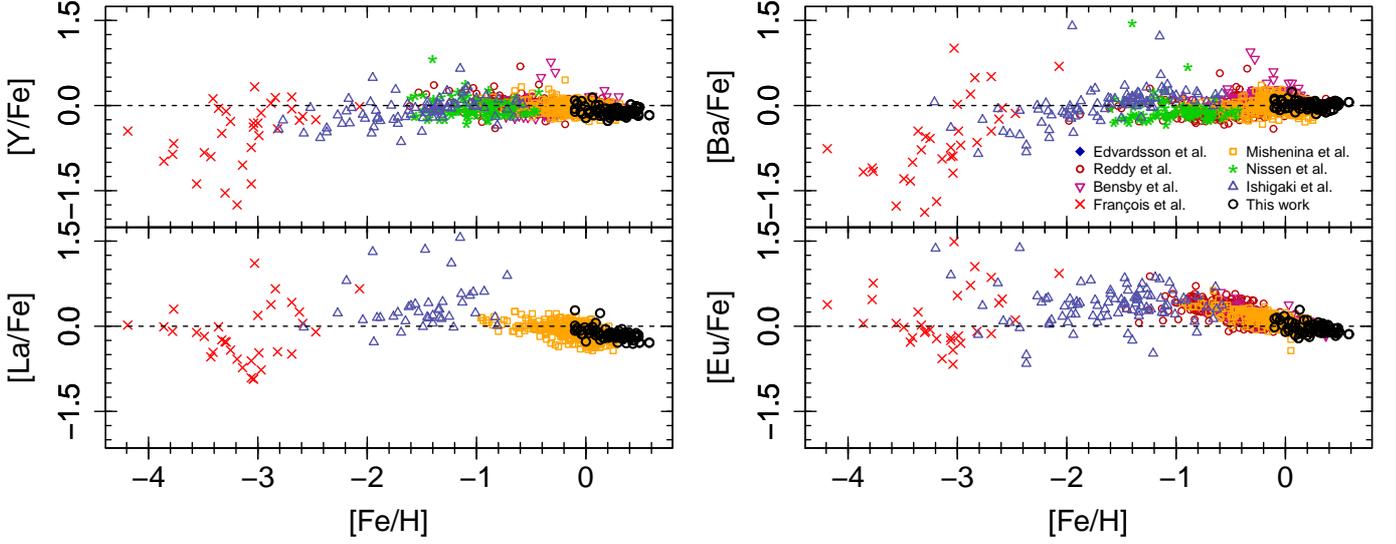}} \\
 \end{tabular}
\caption {Abundances of Y, Ba, La and Eu {\it vs.} metallicities. The sample stars are indicated by 
black open circles. Halo stars studied by \citet{Francois.etal:2007} and 
\citet{Ishigaki.etal:2013} are represented by red crosses and blue triangles, respectively. 
The disk stars are shown as blue diamonds \citep{Edvardsson.etal:1993}, red open circles \citep{Reddy.etal:2003, Reddy.etal:2006},
magenta triangles \citet{Bensby.etal:2005}, green stars \citep{Nissen.Schuster:2011} and yellow squares \citep{Mishenina.etal:2013}.}
\label{Fig_abonds}
\end{figure*}

The final abundances of Y, Ba, La, and Eu are shown in Fig.~\ref{Fig_abonds}, using iron as the reference element. 
We compare our results with data from the literature for halo stars \citep{Francois.etal:2007, Ishigaki.etal:2013, Nissen.Schuster:2011}
and disk stars \citep{Edvardsson.etal:1993, Reddy.etal:2003, Reddy.etal:2006, Bensby.etal:2005,
Nissen.Schuster:2011, Mishenina.etal:2013}. 
The halo data show a large scatter, as pointed out by \citet{Francois.etal:2007}, and
the trend of element-to-Fe ratio for these four elements becomes tighter above [Fe/H]$>$-2.8.
All these samples together show the chemical evolution for these elements. The  
moderate-metallicity halo and thick-disk stars studied by \citet{Ishigaki.etal:2013} show essentially solar ratios, 
except for a few of them, which are enhanced in
La and Ba. There is also a scatter of Eu abundances and a mean overabundance of [Eu/Fe].
The halo and thick-disk data from \citet{Nissen.Schuster:2011} are very homogeneous and close to solar.
The solar-neighbourhood FGK stars from  \citet{Mishenina.etal:2013} illustrate the disk abundance ratios
and metallicities, which reach [Fe/H]$\sim$+0.3. Our sample stars are the most metal-rich sample
and show the abundances of these elements for high metallicities.

%
\begin{figure*}[!ht]
\centering
\begin{tabular}{c}
 \resizebox{\hsize}{!}{\includegraphics[angle=-90]{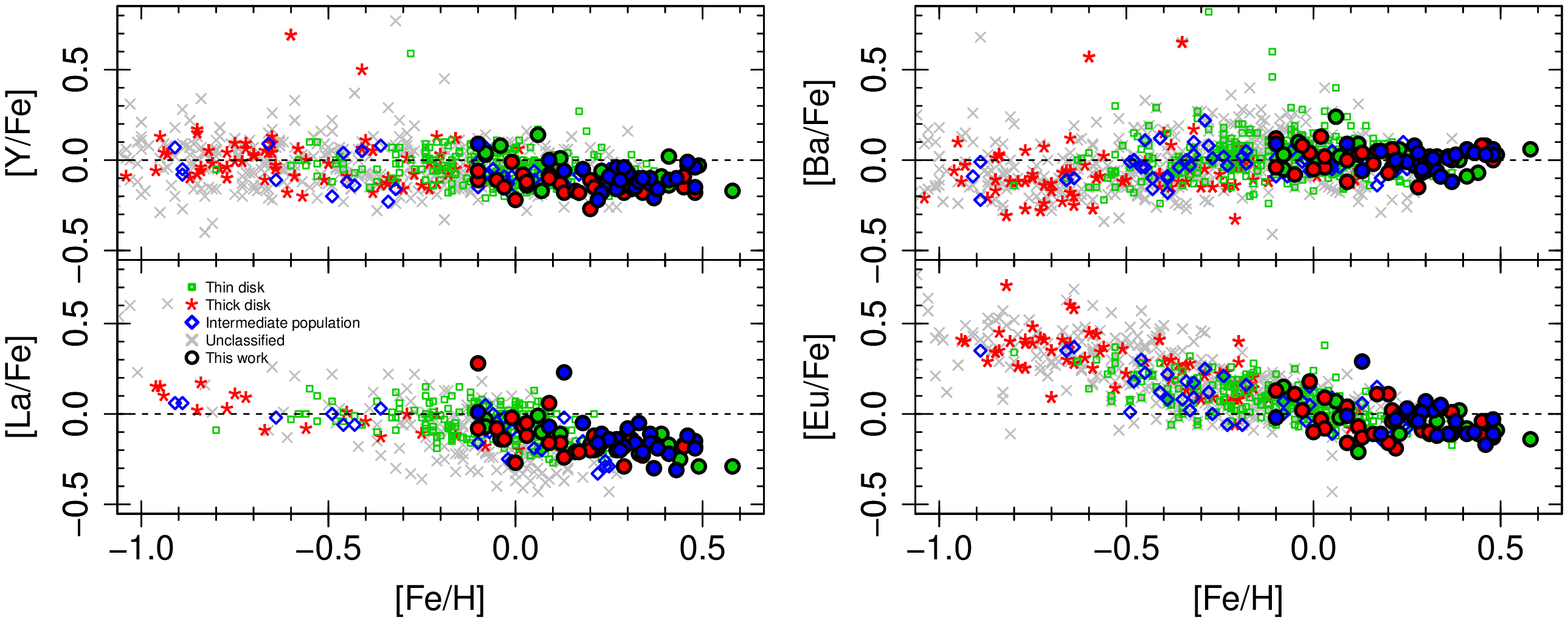}} \\

 \end{tabular}
\caption {Abundances of Y, Ba, La, and Eu {\it vs.} metallicities. The abundances derived in this work (circles)
are compared with the thin- (green squares) and thick-disk (red stars) stars and 
the intermediate population (blue diamonds) from  
\citet{Edvardsson.etal:1993}, \citet{Reddy.etal:2003, Reddy.etal:2006}, 
\citet{Bensby.etal:2005},  \cite{Nissen.Schuster:2011}, \citet{Mishenina.etal:2013}, and \citet{Ishigaki.etal:2013}. 
Grey crosses represent stars with no available U, V, and W velocities, therefore no membership assignment was possible.
The symbols representing our sample stars are filled according to their membership classification: thin disk (green), thick disk (red) and 
intermediate population (blue).}
\label{Fig_abonds2}
\end{figure*}

\subsection{Abundance trends}

In Fig.~\ref{Fig_abonds2} we present a more detailed view of the abundance trends within the Galactic disk by separating the sample 
into thin- and thick-disk components and the intermediate population.
The membership assignment procedure is described in Sect.~\ref{Sec_orbits}. 
Using the U, V, and W space velocities from the Geneva-Copenhagen 
Survey \citep{Holmberg.etal:2009, Casagrande.etal:2011}, the same procedure was 
applied to stars studied by \citet{Edvardsson.etal:1993}, \citet{Reddy.etal:2003, Reddy.etal:2006},
\citet{Bensby.etal:2005}, and \citet{Mishenina.etal:2013}. The assigned stellar population identifications are used
to plot the literature stars with different symbols in Figs. 8-16, according to their kinematic classification
in the terms employed in Paper~I and as briefly described in Sect.~\ref{Sec_orbits}.

The yttrium-to-iron abundance ratio decreases with increasing metallicity, which is compatible with the trend 
found for the disk population. [Ba/Fe] increases with metallicities up to [Fe/H]~$\sim -0.1$, and above this limit, the
[Ba/Fe] ratio  seems to remain constant. Data from the literature indicate solar [La/Fe] values in the metallicity range 
$-0.8 <$~[Fe/H]~$< -0.3$, which
decreases to [La/Fe]~$\sim -0.1$ for stars with [Fe/H]~$\sim 0.0$. Our sample stars folow the trend found for the 
literature data, which presents a decrease of La-to-Fe ratios with metallicity. 
The [La/Fe] drops from [La/Fe]~$\sim -0.1$ for stars with solar iron abundance 
towards [La/Fe]~$\sim -0.2$ at high metallicities.

The literature data show a trend of
Eu-to-Fe ratios dropping from [Eu/Fe]~$\sim$0.4 in the halo and metal-poor thick disk towards
a solar value at the solar metallicity. The [Eu/Fe] values of our sample stars 
follow this trend: they reach slightly sub-solar values ([Eu/Fe]~$\sim -0.1$) 
at high metallicities ([Fe/H]~$> 0.4$). 

%
\begin{figure}[!t]
\centering
\begin{tabular}{c}
 \resizebox{\hsize}{!}{\includegraphics[angle=-90]{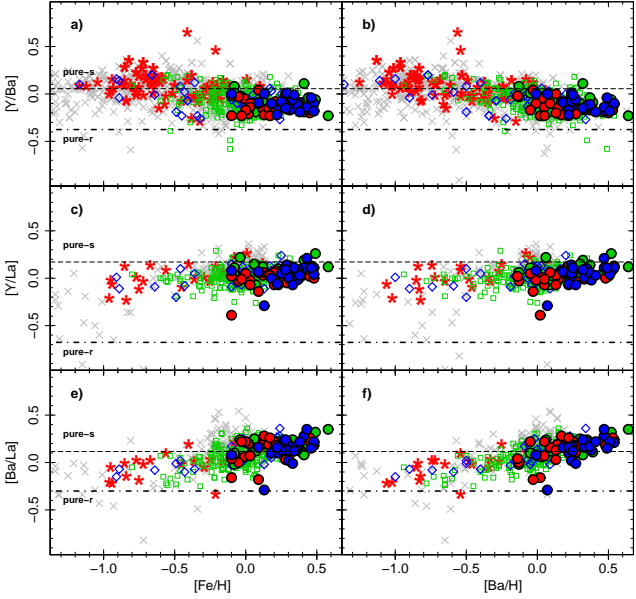}} \\

 \end{tabular}
\caption {Abundance ratios among neutron-capture elements. The symbols are the same as in
Fig.~\ref{Fig_abonds2}. The horizontal lines indicate the abundance ratios for the $r$-process (dash-dotted) and $s$-process (dashed) components
in the solar system, as predicted by \citet{Arlandini.etal:1999}.}
\label{Fig_ratios}
\end{figure}

\subsection{Abundance ratios among neutron-capture elements}

\begin{figure}[!t]
\centering
\begin{tabular}{c}
 \resizebox{\hsize}{!}{\includegraphics[angle=-90]{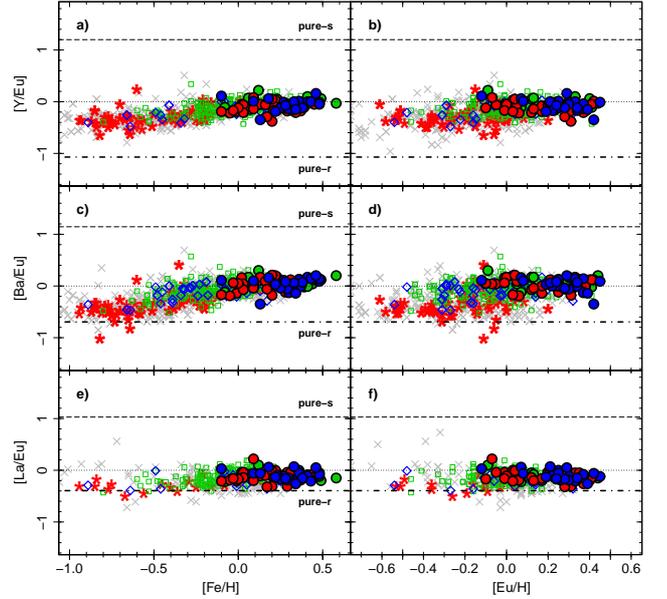}} \\

 \end{tabular}
\caption {Element-to-Eu ratios {\it vs.} stellar metallicities ({\it left}) and europium abundances ({\it right}). 
The symbols are the same as in
Fig.~\ref{Fig_abonds2}.
The horizontal lines indicate the abundance ratios for the $r$-process (dash-dotted) and $s$-process (dashed) components
in the solar system, as predicted by \citet{Arlandini.etal:1999}.}
\label{Fig_XEu_ratio}
\end{figure}

Abundance ratios among neutron-capture elements can provide clues on possible
nucleosynthesis sites of these elements. For this purpose, we inspect the
abundance ratios among Y, Ba, and La (Fig.~\ref{Fig_ratios}), as well as the ratio 
between the abundances of these elements and the $r$-element Eu abundances (Fig.~\ref{Fig_XEu_ratio}).

The ratio of heavy (second peak) to light (first peak) $s$-process elements
depends on the metallicity, as discussed by \citet{Gallino.etal:1998} and \citet{Busso.etal:1999}. 
In Fig.~\ref{Fig_ratios}, we show the [Y/Ba], [Y/La] and [Ba/La] vs. [Fe/H] and vs. [Ba/H]. 
Stars with abundance ratios between the pure-$s$ and pure-$r$ lines should result from both processes. 
[Y/Ba] (Fig.~\ref{Fig_ratios}a, b)  is closer to the pure-$s$ line and tends to be compatible with the s-process production.
It is slightly enhanced for stars with ${\rm [Fe/H]} \lesssim -0.5$, which maybe requires an additional nucleosynthesis process
as suggested by \citet{Francois.etal:2007} and  \cite{Ishigaki.etal:2013}, and [Y/Ba]~$\lesssim$~0.0 for metallicities above this limit.

The Y-to-La ratios (Figs. \ref{Fig_ratios}c,d) are approximately constant in stars with metallicities ranging from
$-1.0$ to $+0.3$. The most metal-rich stars in our sample ([Fe/H]~$> +0.4$) appear to have slightly higher [Y/La] values, with 
values closer to the pure-$s$ line. The [Ba/La] values (Figs. \ref{Fig_ratios}e,f) increase with [Fe/H], 
from [Ba/La]~$\sim -0.1$ for metallicity $-1.0$ up 
to [Ba/La]~$\sim +0.2$ for high metallicities.

[Ba/Eu] vs. [Fe/H] and vs. [Eu/H] values (Figs. \ref{Fig_XEu_ratio}c,d) are low for thick-disk stars at low metallicities, 
indicating that these stars were probably enriched by SNe type II, at early times in the Galaxy.  
[Ba/Eu] increases with [Fe/H] and reaches solar values around [Fe/H]~$\sim 0.0$.
[Y/Eu] and [La/Eu] vs. [Fe/H] and [Eu/H] (Figs. \ref{Fig_XEu_ratio}a,b,e,f) are low for metal-poor thick disk stars,
and essentially solar for most of the sample thin- and thick-disk stars and intermediate populations. 

%

\subsection{Neutron-capture versus $\alpha$-elements}

\begin{figure*}[!ht]
\centering
\begin{tabular}{c}
 \resizebox{0.9\hsize}{!}{\includegraphics[angle=-90]{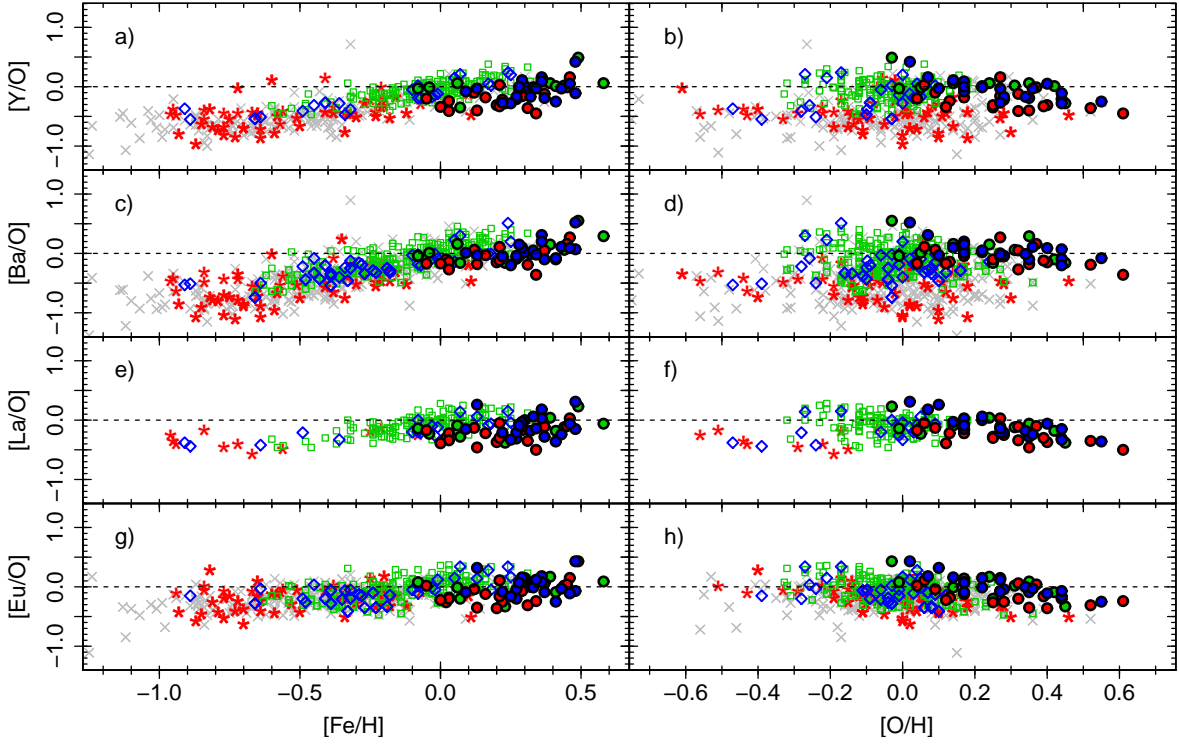}} \\

 \end{tabular}
\caption {Abundance ratios between neutron-capture elements and oxygen as a function of metallicity ({\it left})
and [O/H] ({\it right}). The symbols are the same as in
Fig.~\ref{Fig_abonds2}.}
\label{Fig_XO_ratios}
\end{figure*}

\begin{figure*}[!ht]
\centering
\begin{tabular}{c}
 \resizebox{0.9\hsize}{!}{\includegraphics[angle=-90]{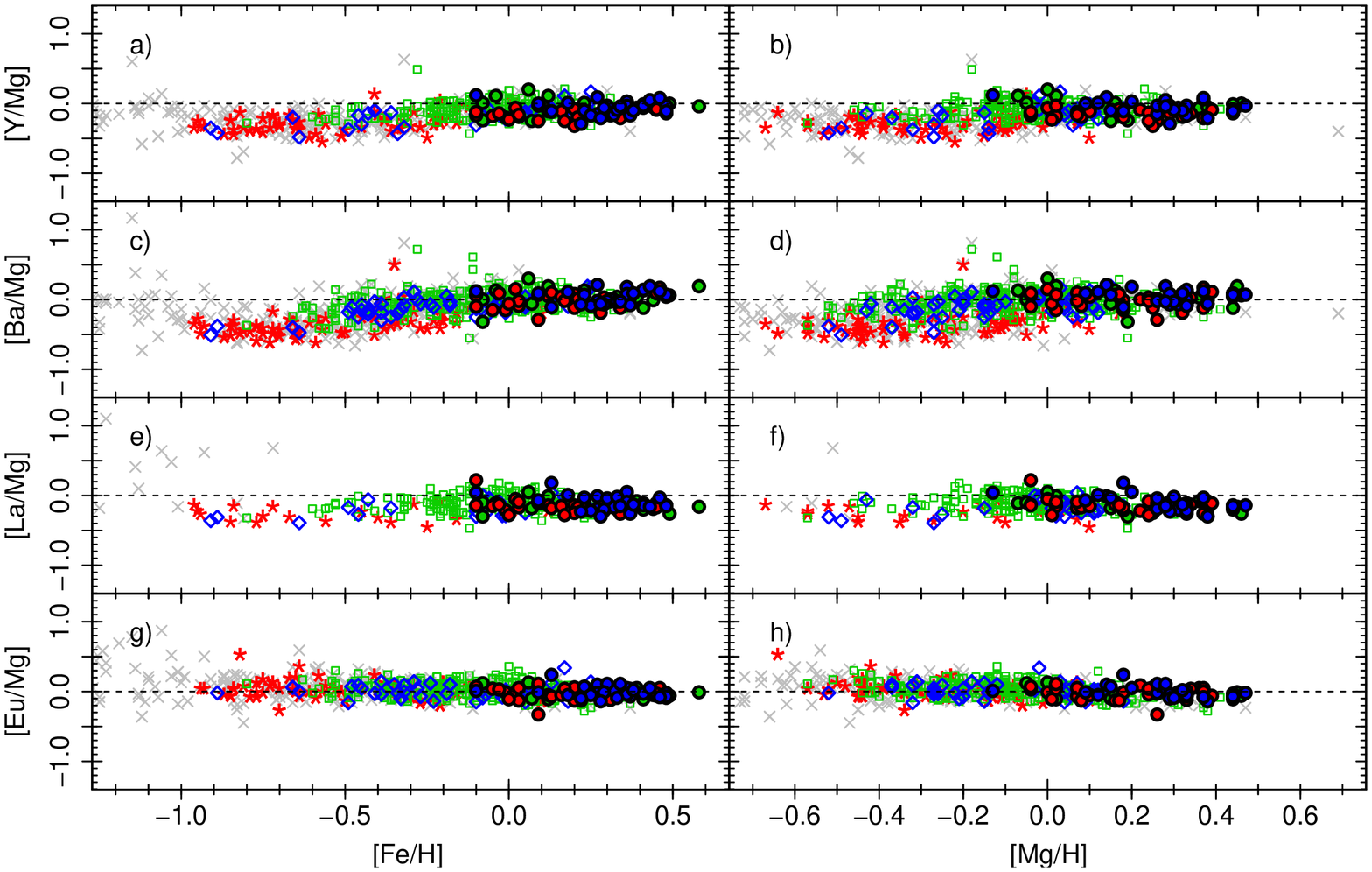}} \\

 \end{tabular}
\caption {Abundance ratios between neutron-capture elements and magnesium as a function of metallicity ({\it left})
and [Mg/H] ({\it right}). The symbols are the same as in
Fig.~\ref{Fig_abonds2}.}
\label{Fig_XMg_ratios}
\end{figure*}

Another interesting piece of information comes from inspecting the abundance
ratios between heavy and $\alpha$-elements. 
Iron, which is frequently used as the reference element, is produced in 
both SNe II and SNe Ia events. The $r$-process component of the heavy elements and the 
 $\alpha$-elements, in particular oxygen and magnesium, are produced in SNe II. Therefore,
it is interesting to investigate how the
abundance ratios of neutron-capture elements vary as a function of the abundances of oxygen and magnesium, as shown in 
Figs. \ref{Fig_XO_ratios} and \ref{Fig_XMg_ratios}.

%

%

%
\begin{figure*}[!ht]
\centering
\begin{tabular}{c}
  \resizebox{0.9\hsize}{!}{\includegraphics[angle=-90]{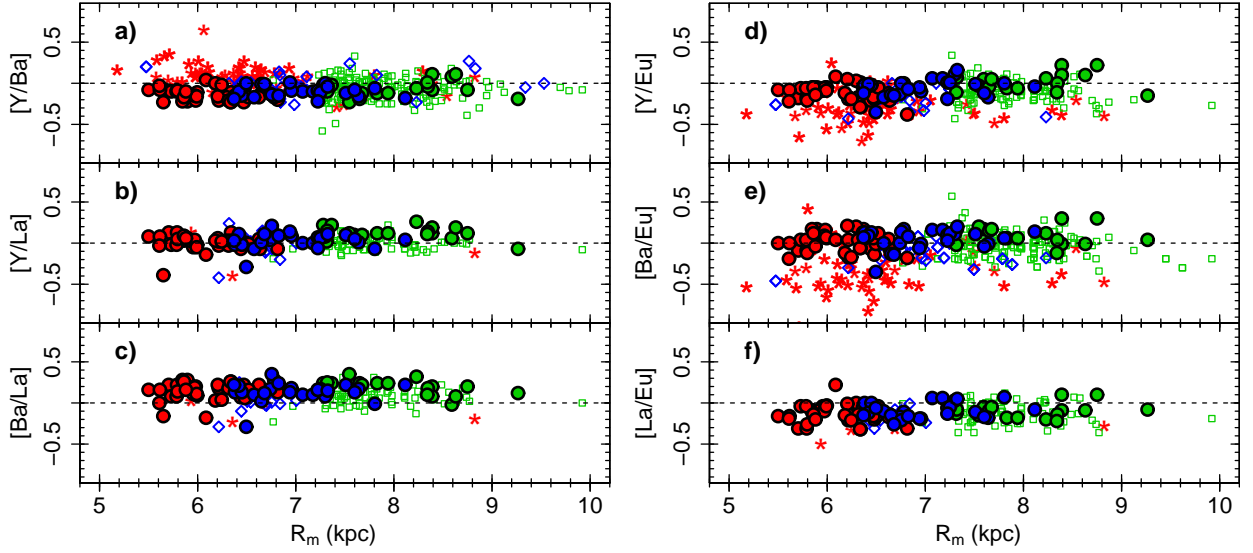}} \\

 \end{tabular}
\caption {Abundance ratios among neutron-capture element {\it vs.} mean Galactocentric distance. The symbols are the same as in
Fig.~\ref{Fig_abonds2}.}
\label{Fig_ratios_Rm}
\end{figure*}

%
\begin{figure*}[!ht]
\centering
  \resizebox{0.9\hsize}{!}{\includegraphics[angle=-90]{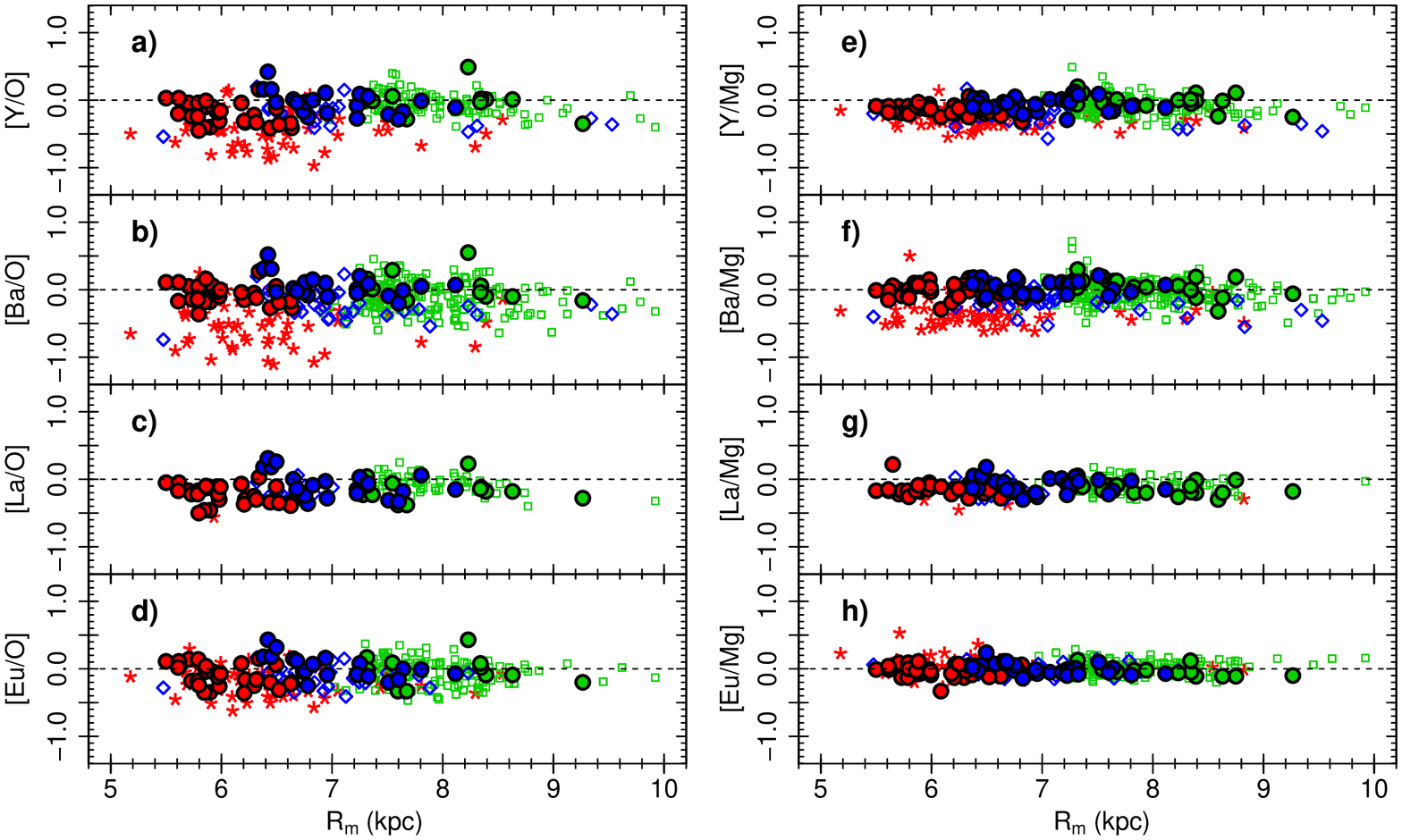}} \\

\caption {[Element/O] ({\it left panels}) and [element/Mg] ({\it right panels}) ratios {\it vs.} mean Galactocentric distances. The symbols are the same as
in Fig.~\ref{Fig_abonds2}.}
\label{Fig_ratios_alpha_Rm}
\end{figure*}

[Y/O] and [Ba/O] vs. [Fe/H] (Figs. \ref{Fig_XO_ratios}a,c) increase with metallicity for all populations,
but our sample stars show lower [Y,Ba,La/O] with respect to the thin disk.
It is interesting to note that La and Eu have similar behaviours: [La/O] and [Eu/O] vs. [Fe/H] (Figs. \ref{Fig_XO_ratios}e,g)
increase steadily with metallicity, and our results are compatible with the thin-disk behaviour.
More interesting in these comparisons is the behaviour of [Y/O] and [Ba/O] vs. [O/H] (Figs. \ref{Fig_XO_ratios}b,d),
which indicates that our sample has higher oxygen abundances than the thin-disk stars,
and show higher [Y,Ba/O] ratios relative to literature thick-disk stars.  The same applies to  
[La/O] and [Eu/O] vs. [O/H] (Figs. \ref{Fig_XO_ratios}f,h), but in this case, the literature thick-disk stars have
the same behaviour as the sample stars.

%
\begin{figure*}[!ht]
\centering
\begin{tabular}{ccc}
  \resizebox{0.3\hsize}{!}{\includegraphics{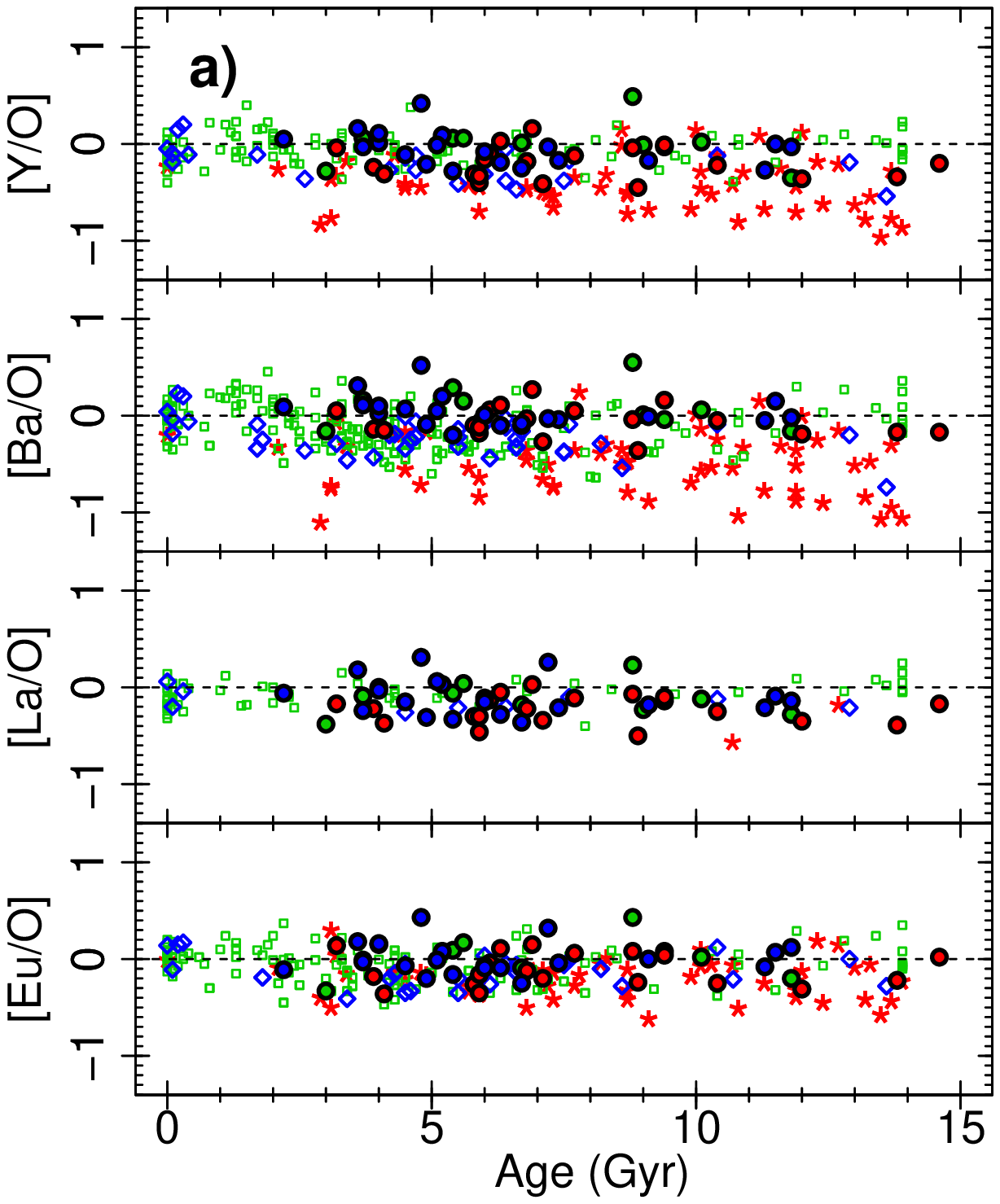}} &
  \resizebox{0.3\hsize}{!}{\includegraphics{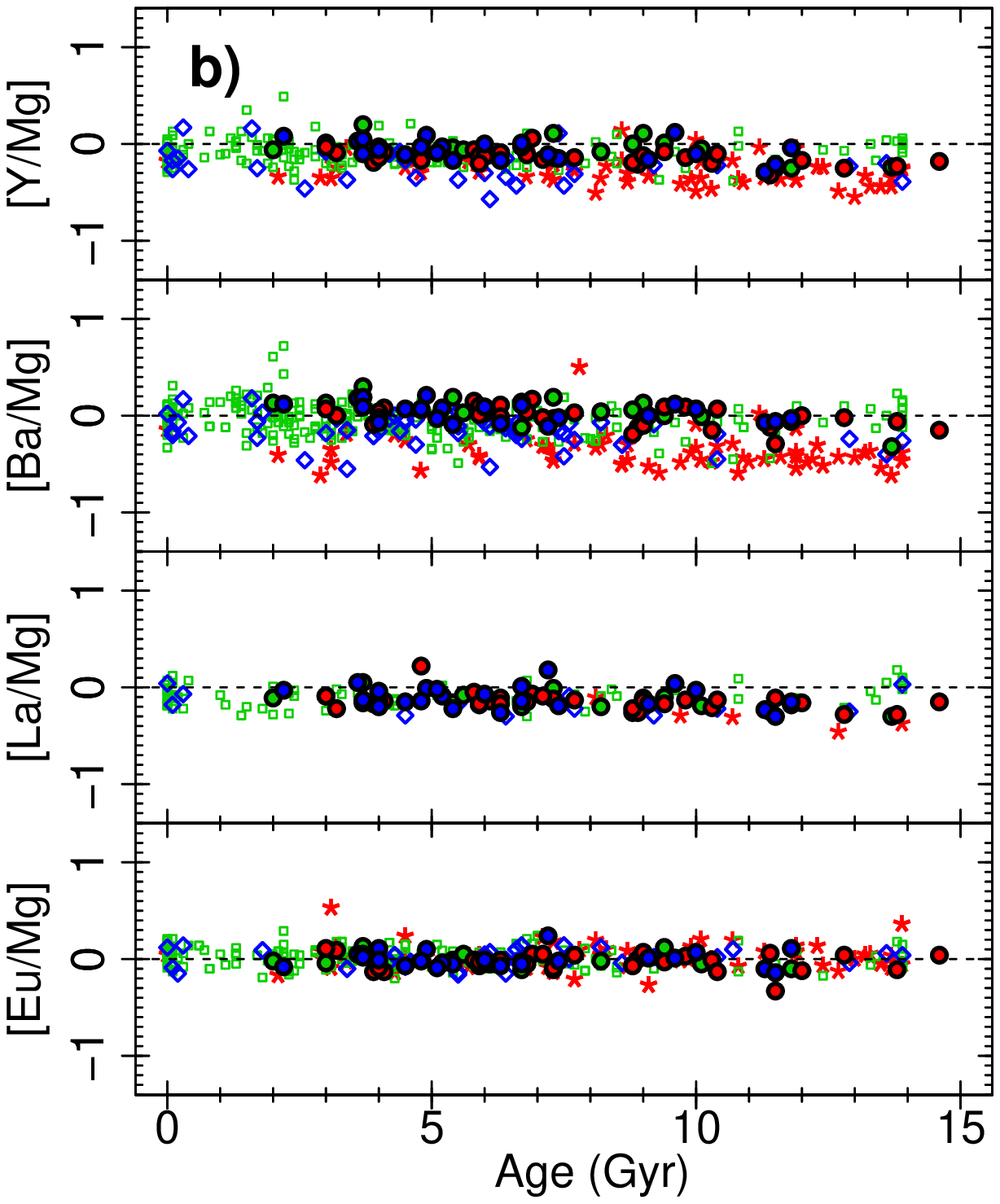}} &
    \resizebox{0.3\hsize}{!}{\includegraphics{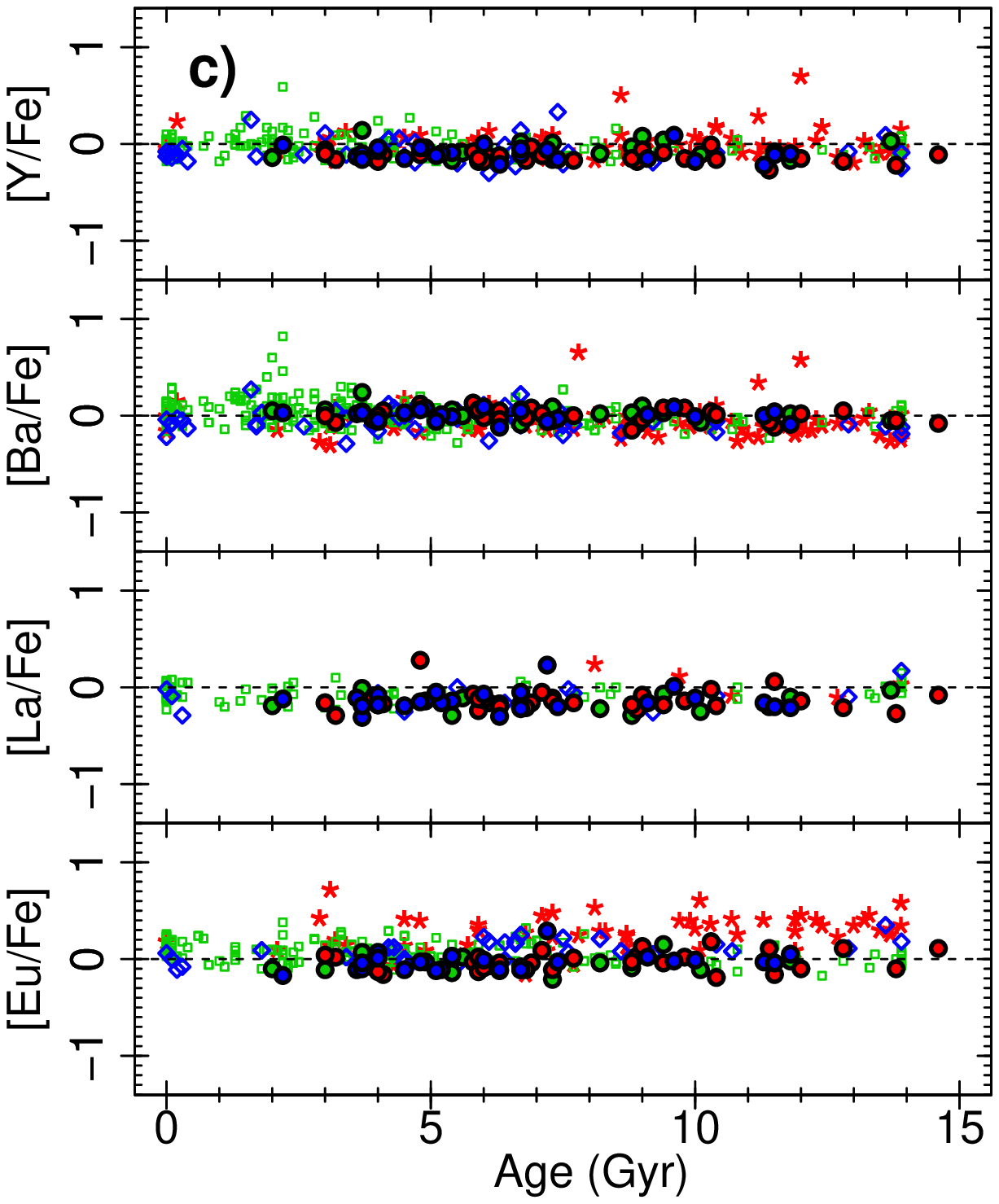}} \\

 \end{tabular}
\caption {Element-to-oxygen ({\it left}), element-to-magnesium ({\it middle}), and element-to-iron ratios ({\it right}) {\it vs.} stellar ages.
The notation is the same as in
Fig.~\ref{Fig_abonds2}. Ages for the metal-rich sample were derived as described in Paper~I.
 Ages for samples from the literature were retrieved from the catalogue of \citet{Casagrande.etal:2011} .}
\label{Fig_abonds_Age}
\end{figure*}

%
\begin{figure}[!ht]
\centering
\begin{tabular}{c}
 \resizebox{0.95\hsize}{!}{\includegraphics{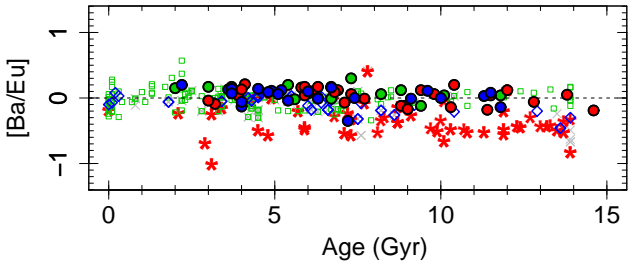}} \\

 \end{tabular}
\caption {[Ba/Eu] {\it vs.} stellar ages.The notation is the same as in
Fig.~\ref{Fig_abonds2}.}
\label{Fig_EuBa_Age}
\end{figure}

Figure~\ref{Fig_XMg_ratios} shows [Y,Ba,La,Eu/Mg] vs. [Fe/H] and [Mg/H]. It is similar to 
 Fig.~\ref{Fig_XO_ratios}, but it shows a strikingly better defined
chemical enrichment. 
Besides, [Y,Ba/Mg] ratios are compatible with the ratios in thin-disk stars.
[Eu/Mg] is constant across the whole metallicity range, indicating that Eu and Mg are produced in constant proportions
in SNe type II. 
It is important to note that for metal-rich stars, magnesium might be a better reference element
than oxygen. A plot of [Mg/O] vs. [Mg/H],  where oxygen drops relative to magnesium at high metallicities, was given for example in
 \citet{Bensby.etal:2004}.

\subsection{Abundances vs. Galactocentric distances}

Figures \ref{Fig_ratios_Rm} and \ref{Fig_ratios_alpha_Rm} show abundance ratios vs.
 the mean Galactocentric distances, defined as $R_{\rm m}$ = (R$_{\rm a}$ + R$_{\rm p}$)/2, 
that is the mean of the apocentric and pericentric distances. 
The abundance ratios in our sample stars appear to be constant with $R_{\rm m}$. 
On the other hand, data from the literature indicate that [Y,Ba/Eu], [Y,Ba,Eu/O], and [Y,Ba/Mg] increase 
with $R_{\rm m}$. 
The observed increase of [Y,Ba/Eu] with $R_{\rm m}$ could mean that 
stars close to the centre are older than the outermost stars. Note that Ba has been used as an age indicator by
 \citet{Edvardsson.etal:1993}.

 [Y/Eu,O,Mg] and  [Ba/Eu,O,Mg] (Figs. \ref{Fig_ratios_Rm}d,e and \ref{Fig_ratios_alpha_Rm}a,b,e,f) for our thick-disk stars are
 clearly enhanced relative to the literature thick-disk stars.
 These ratios  are more compatible to the literature thin-disk stars,
 but they are located in the inner regions, and might indicate 
 that our sample of stars could be old inner thin-disk stars (cf. Paper~I).

\subsection{Abundances vs. stellar ages}

The chemical evolution of our Galaxy with time can be inspected based on estimate of stellar ages. 
 However, age is difficult to derive for most  stars. 
 In Paper~I we derived stellar ages with uncertainties smaller than 30\% for 36 sample stars.

 Figures \ref{Fig_abonds_Age}a,b,c present [Y,Ba,La,Eu/O], [Y,Ba,La,Eu/Mg], [Y,Ba,La,Eu/Fe] vs. age,
respectively.
 There is a clear distinction between the literature thin- and thick-disk stars in the [Y/O], [Ba/O] vs. ages plane. 
 Our sample matches the thin-disk stars, again indicating that they might be old thin-disk stars. 
 The scatter also increases with age, and different ratios might indicate that stars were formed at 
 different Galactocentric distances. 
 [La/O] is more similar to  [Eu/O], and both show solar values at all ages, except that
[Eu/O] for thick-disk stars is somewhat lower than in thin-disk stars.
 
Figure \ref{Fig_abonds_Age}b is similar to Fig.~\ref{Fig_abonds_Age}a, but with a smaller scatter. 
 Abundances ratios using Fe as the reference element (Fig.~\ref{Fig_abonds_Age}c) are constant with stellar age for all populations, 
except [Eu/Fe] in thick-disk stars, which increases with age.
Thus the plots relative to O are better suited to distinguish different stellar populations. 
It is interesting to note that, again, the Eu and Mg production is tighly correlated
 for all ages. 
 
 [Ba/Eu] ratio, which is plotted against stellar ages in Fig. \ref{Fig_EuBa_Age}, is constant and close to solar 
 across the whole age range for thin-disk and our sample stars. On the other hand, thick-disk stars show 
 low Ba-to-Eu ratios.

\subsection{Comparison with bulge stars}

We compared the abundances of the sample stars and the bulge microlensed dwarf and subgiant stars 
studied by \citet{Bensby.etal:2010b, Bensby.etal:2011, Bensby.etal:2013}, and red giants from the
bulge Plaut field by \citet{Johnson.etal:2012, Johnson.etal:2013}.

\begin{figure*}[!ht]
\centering
\begin{tabular}{c}
  \resizebox{0.95\hsize}{!}{\includegraphics[angle=-90]{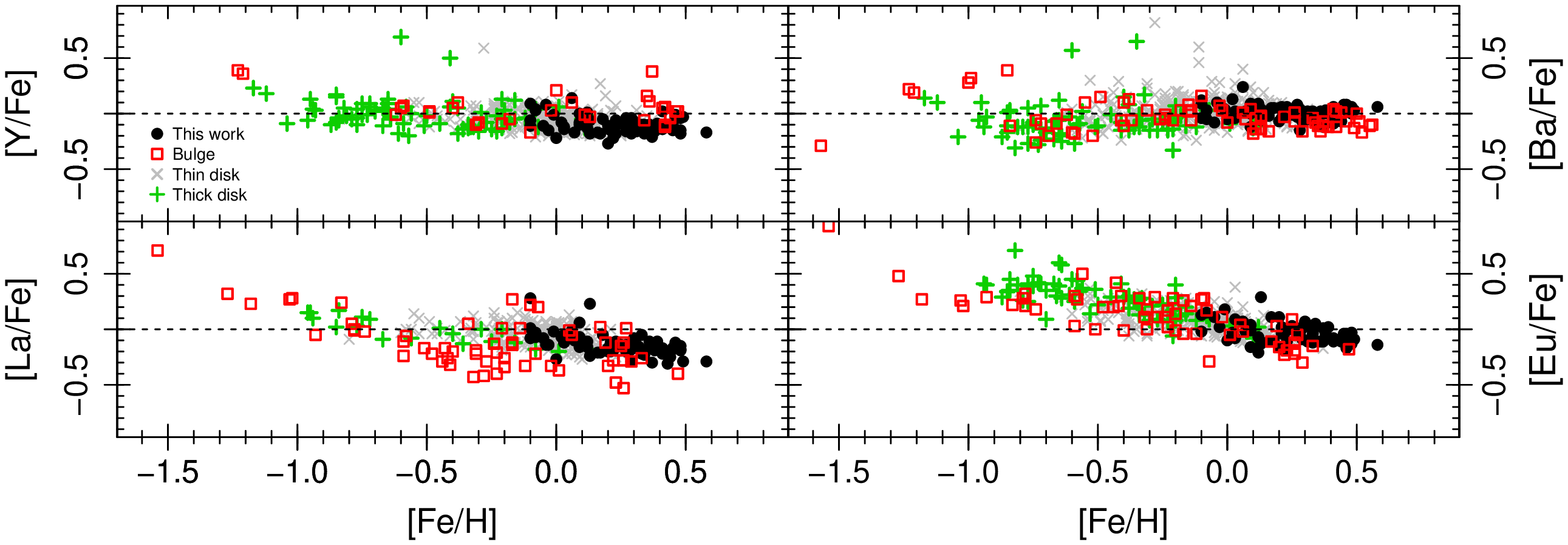}} \\
 \end{tabular}
\caption {Abundances of Y, Ba, La, and Eu vs. [Fe/H] for the sample stars (black dots) and bulge stars (red squares). 
The abundances of the bulge population were derived by \citet[][microlensed dwarf stars]{Bensby.etal:2010b, Bensby.etal:2011, Bensby.etal:2013} 
  and by \citet[][RGB stars]{Johnson.etal:2012}. The grey and green crosses
  indicate thin- and thick-disk stars. }
\label{Fig_bulge_YBaFe}
\end{figure*}

\begin{figure*}[!ht]
\centering
\begin{tabular}{c}
\vspace{-1.2cm}
  \resizebox{0.95\hsize}{!}{\includegraphics[angle=-90]{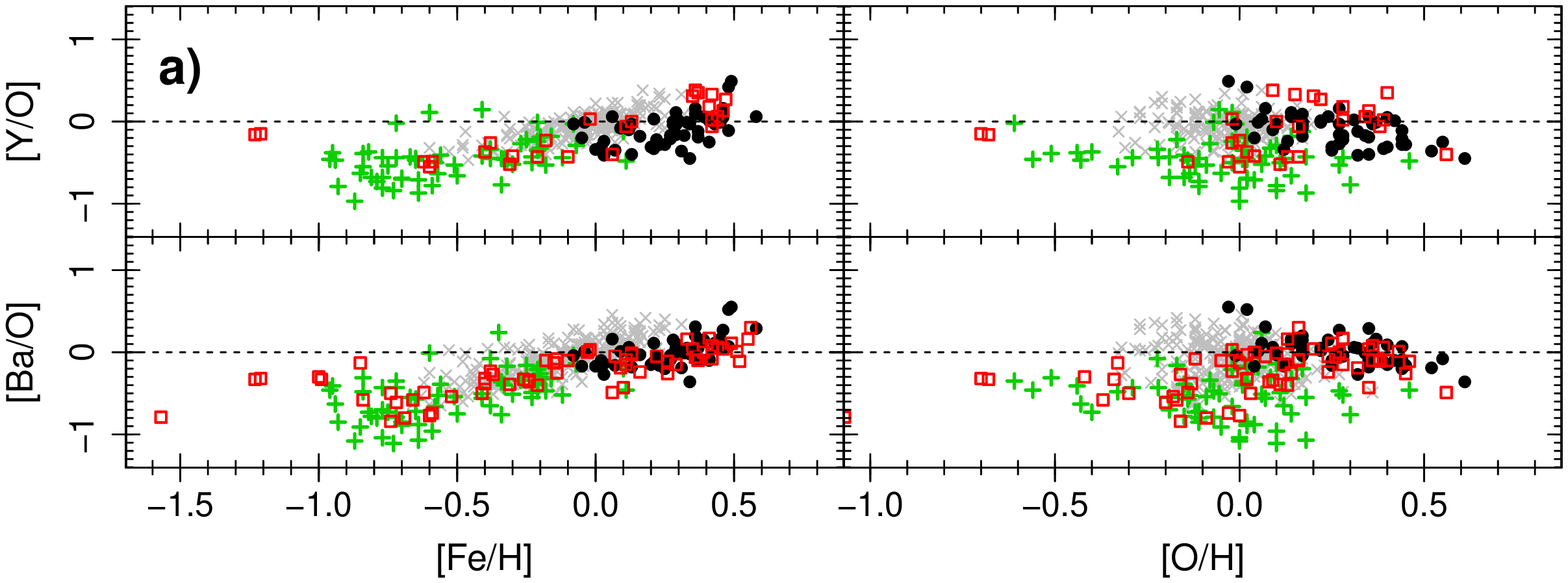}} \\
  \resizebox{0.95\hsize}{!}{\includegraphics[angle=-90]{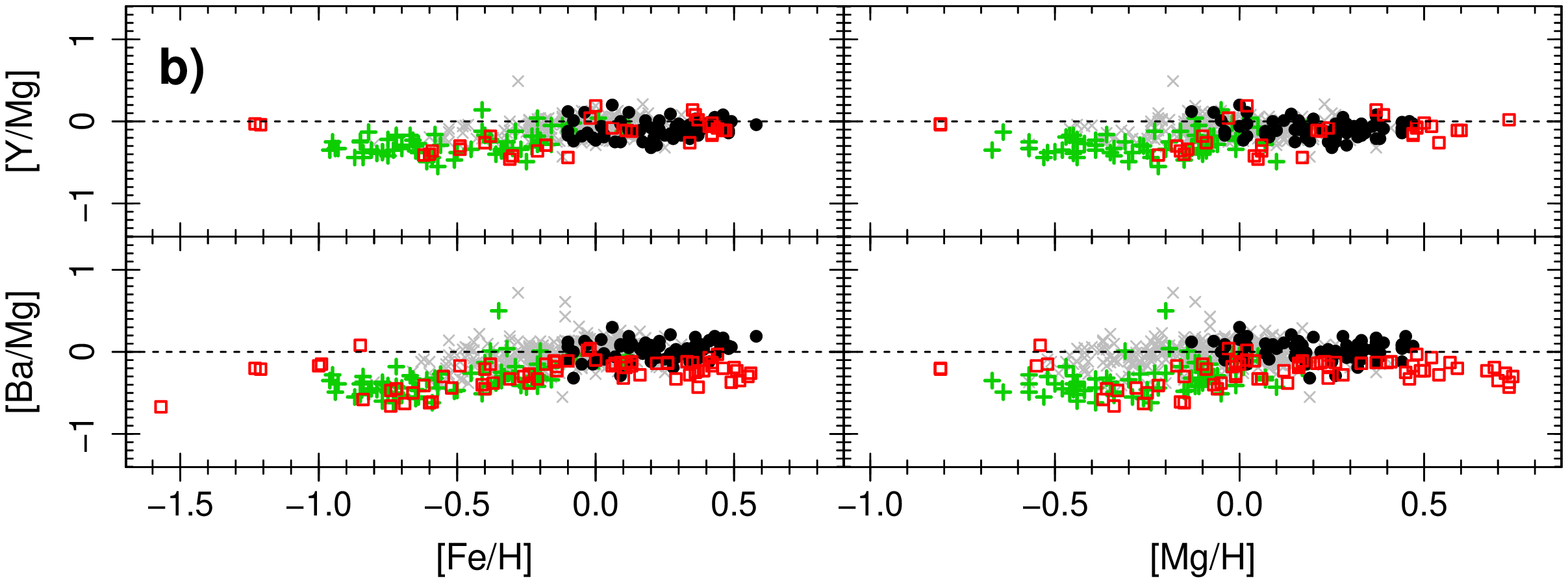}} \\
 \end{tabular}
\caption {Abundance ratios between neutron-capture elements and $\alpha$-elements for the sample stars and for bulge stars. 
  {\it Upper panel:} [Y/O] and [Ba/O] abundance ratios vs. [Fe/H] ({\it left}) and [O/H] ({\it right}). 
  {\it Lower panel:} [Y/Mg] and [Ba/Mg] abundance ratios vs. [Fe/H] ({\it left}) and [Mg/H] ({\it right}). 
  The notation is the same as in Fig.~\ref{Fig_bulge_YBaFe}.}
\label{Fig_bulge_YBaOMg}
\end{figure*}

Figure \ref{Fig_bulge_YBaFe} presents Y, Ba, La, and Eu abundances as a function of [Fe/H].
 Literature thin- and thick-disk data are shown for comparison. 
There is no clear distinction between the bulge stars and our sample 
in terms of metallicity and [Y,Ba,Eu/Fe] abundance ratios. 
Figure~\ref{Fig_bulge_YBaOMg} shows the abundance of Y and Ba using oxygen and magnesium as the reference element.
There is a clear distinction between the trends in thin-disk and bulge stars, whereas our 
sample follows the trend observed for the bulge. 
In addition, these abundance ratios indicate that the sample stars may be the high-metallicity extension of the thick-disk stars. 
The trend of the thick-disk stars together with our sample is similar to the trend observed for the bulge stars.

%

%

\section{Discussion}
\label{Sec_discussion}

The mean Galactocentric distance is related to thin- and thick-disk membership probability, 
as can be seen in Figs. \ref{Fig_ratios_Rm} and
\ref{Fig_ratios_alpha_Rm}. 
The thick-disk stars 
have the lowest mean distances  $R_{\rm m}$, since they have pericentric distances closer to the Galactic centre than thin-disk stars. 
For this reason, it is difficult to separate 
the inner thin-disk population and the thick-disk components based on kinematics alone.

Abundance ratios  reflect the enrichment of the interstellar medium at the time the star was formed, 
and depends on the region of the Galaxy where
the star formation occurred  (e.g. radial gradients, that is, inner parts of the Galaxy contain stars with higher metallicities 
than the outer parts). 
If there is much radial migration in the dynamical evolution of the disk, interpreting [X/Fe] vs. [Fe/H]
for stars in the solar neighbourhood is difficult. 
However, while the kinematics of the stars may change during their lifetime, the stellar abundances do not. 
Therefore, chemical tagging is more efficient in identifying different stellar populations.

Our sample shows clear similarities with respect to the thin-disk stars, but with a few indicators
point a different nucleosynthesis for these two stellar populations. 
These differences are, in particular, better seen when oxygen is used as the reference element,
for instance as [Y/O] and [Ba/O] vs. [Fe/H] and [O/H] (Figs.~\ref{Fig_XO_ratios}a,b,c,d).
Our sample stars show lower [Y/O] and [Ba/O] ratios, higher oxygen abundances,  
and lower [Mg/O] than thin-disk stars.

Figure \ref{Fig_bulge_YBaOMg}, which shows [Y/O] and [Ba/O] vs. [Fe/H] and [O/H],
and [Y/Mg] and [Ba/Mg] vs. [Fe/H] and [Mg/H], suggests similarities between our sample and bulge stars. 
The lower [Y/O] and [Ba/O] vs. [O/H] relative  to thin-disk stars may be due to the older ages of our sample,
since the contribution from AGB stars to the $s$-element abundances is expected to be lower in the older stars.

\section{Summary}
\label{Sec_summary}

We have derived the abundances of the heavy elements Y, Ba, La, and Eu for the sample of 71 metal-rich
stars for which a detailed analysis was carried out in \citet{Trevisan.etal:2011}.

Chemical tagging is a most efficient indicator for identifying different stellar populations.
 We compared our results with data from the literature for samples of thin- and thick-disk stars (Fig.~\ref{Fig_abonds2}).
Our sample of stars is more metal-rich  than these samples.
 We inspected the abundance ratios among neutron-capture elements (Figs.~\ref{Fig_ratios} and \ref{Fig_XEu_ratio}) and 
between heavy and $\alpha$-elements (Figs. \ref{Fig_XO_ratios} and \ref{Fig_XMg_ratios}).
Our sample shows clear similarities with the thin-disk stars,
but there are differences, in particular, the [Y/O] and [Ba/O] vs. [Fe/H] and [O/H]
behaviour (Fig.~\ref{Fig_XO_ratios}a,b,c,d).

A comparison with the bulge abundances showed that [Y/Fe,O,Mg] and [Ba/Fe,O,Mg] vs. [Fe/H], [O/H], and [Mg/H]
 (Figs. \ref{Fig_bulge_YBaFe} and \ref{Fig_bulge_YBaOMg})
 suggest similarities between our sample and bulge stars.
It would be important to be able to study more bulge stars in terms of the discriminators
La, and Eu vs. the alpha-elements O and Mg, and compare them with our sample
 to verify the degree of similarity between this sample and bulge stars.

The similar behaviour of La and Eu, which seems to vary in lockstep,
as well as  the
constant value of [Eu/Mg] vs. [Mg/H] is remarkable. It suggests that Eu and Mg
are produced in the same SNe type II,
in the same proportions, at least at metallicities [Fe/H]~$> -1$.

In conclusion, our results indicate that the sample stars might either be bulge stars,
which justifies the denomination ``bulge-like'', as previously suggested by \citet{Pompeia.etal:2003}, 
or inner old thin-disk stars \citep{Haywood:2008}. In either cases they would have been
ejected by the bar towards the solar neighbourhood.  It is less likely that they represent a high-metallicity complement to
the thick-disk stars. Finally, since some abundance ratios are different from those in the bulge and thin-
and thick-disk stars, the sample stars may be yet another stellar population. This makes additional verifications desirable.

\section*{Acknowledgments}

We would like to thank the anonymous referee for his or her helpful 
comments, which helped to improve this manuscript.
MT acknowledges the support of FAPESP, process no. 2012/05142-5. 
The observations were carried out within Brazilian time in a ESO-ON agreement, 
and within an IAG-ON agreement funded by FAPESP project n$^{\circ}$ 1998/10138-8.
We acknowledge partial financial support from CNPq and Fapesp.

\onecolumn
\normalsize

\begin{longtable}{lcrrrrr}
\caption{Final abundances}\\
\hline\hline

Star    & $T_{\rm eff}$ (K) & \multicolumn{1}{c}{[Fe/H]} & \multicolumn{1}{c}{[Y/Fe]} & \multicolumn{1}{c}{[Ba/Fe]} & \multicolumn{1}{c}{[La/Fe]} & \multicolumn{1}{c}{[Eu/Fe]}\\

\hline
\endfirsthead
\caption{continued.} \\
\hline\hline

Star    & $T_{\rm eff}$ (K) & \multicolumn{1}{c}{[Fe/H]} & \multicolumn{1}{c}{[Y/Fe]} & \multicolumn{1}{c}{[Ba/Fe]} & \multicolumn{1}{c}{[La/Fe]} & \multicolumn{1}{c}{[Eu/Fe]}\\

\hline
\endhead
\hline
\multicolumn{7}{l}{The number of lines used to compute the final abundances are indicated between parenthesis.} \\
\multicolumn{7}{l}{The errors correspond to line-to-line scatter.}
\endfoot

G 161-029    & 4798 &  0.01 $\pm$ 0.08 &  0.01 $\pm$  --- (1) & -0.02 $\pm$ 0.09 (4) & -0.07 $\pm$ 0.27 (2) & \multicolumn{1}{c}{---} \\
BD-02   180  & 5004 &  0.33 $\pm$ 0.04 & -0.22 $\pm$ 0.05 (3) &  0.05 $\pm$ 0.10 (4) & -0.07 $\pm$ 0.08 (2) &  0.00 $\pm$ 0.14 (2) \\ 
BD-05  5798  & 4902 &  0.20 $\pm$ 0.05 & -0.27 $\pm$ 0.07 (3) & -0.07 $\pm$ 0.13 (4) & -0.20 $\pm$ 0.19 (2) &  0.11 $\pm$  --- (1) \\	
BD-17  6035  & 4892 &  0.09 $\pm$ 0.05 & -0.08 $\pm$ 0.07 (3) & -0.12 $\pm$ 0.12 (3) &  0.06 $\pm$  --- (1) & -0.16 $\pm$  --- (1) \\	
CD-32  0327  & 4957 & -0.01 $\pm$ 0.05 & -0.01 $\pm$ 0.05 (2) &  0.04 $\pm$ 0.07 (4) & -0.02 $\pm$ 0.01 (2) &  0.18 $\pm$ 0.09 (2) \\	
CD-40 15036  & 5429 & -0.03 $\pm$ 0.07 & -0.15 $\pm$ 0.08 (4) &  0.08 $\pm$ 0.09 (4) & -0.14 $\pm$ 0.02 (2) &  0.02 $\pm$  --- (1) \\	
HD     8389  & 5274 &  0.58 $\pm$ 0.04 & -0.17 $\pm$ 0.10 (5) &  0.06 $\pm$ 0.13 (4) & -0.29 $\pm$ 0.13 (4) & -0.14 $\pm$  --- (1) \\	
HD     9174  & 5599 &  0.41 $\pm$ 0.06 & -0.11 $\pm$ 0.06 (5) &  0.06 $\pm$ 0.09 (4) & -0.22 $\pm$ 0.08 (3) & -0.11 $\pm$ 0.10 (2) \\	
HD     9424  & 5449 &  0.12 $\pm$ 0.05 & -0.13 $\pm$ 0.06 (4) &  0.03 $\pm$ 0.04 (4) & -0.16 $\pm$ 0.15 (2) & -0.07 $\pm$  --- (1) \\	  
HD    10576  & 5929 &  0.02 $\pm$ 0.06 & -0.08 $\pm$ 0.04 (5) &  0.13 $\pm$ 0.06 (4) & -0.07 $\pm$ 0.12 (3) & -0.03 $\pm$ 0.28 (2) \\  
HD    11608  & 4966 &  0.39 $\pm$ 0.05 & -0.09 $\pm$ 0.10 (3) &  0.00 $\pm$ 0.11 (4) & -0.11 $\pm$ 0.07 (3) &  0.02 $\pm$  --- (1) \\  
HD    12789  & 5810 &  0.27 $\pm$ 0.05 & -0.04 $\pm$ 0.09 (5) &  0.08 $\pm$ 0.09 (4) & -0.14 $\pm$ 0.12 (4) & -0.03 $\pm$ 0.00 (2) \\  
HD    13386  & 5269 &  0.36 $\pm$ 0.04 & -0.13 $\pm$ 0.09 (5) &  0.02 $\pm$ 0.06 (4) & -0.11 $\pm$ 0.06 (3) & -0.11 $\pm$ 0.10 (2) \\  
HD    15133  & 5223 &  0.46 $\pm$ 0.04 & -0.03 $\pm$ 0.09 (5) &  0.08 $\pm$ 0.10 (4) & -0.16 $\pm$ 0.08 (4) & -0.04 $\pm$ 0.04 (2) \\  
HD    15555  & 4867 &  0.37 $\pm$ 0.05 & -0.21 $\pm$ 0.03 (5) & -0.12 $\pm$ 0.06 (4) & -0.30 $\pm$ 0.08 (3) & -0.11 $\pm$ 0.01 (2) \\  
HD    16905  & 4866 &  0.27 $\pm$ 0.04 & -0.16 $\pm$ 0.08 (2) & -0.03 $\pm$ 0.09 (4) & -0.20 $\pm$ 0.04 (3) & -0.03 $\pm$ 0.14 (2) \\  
HD    25061  & 5307 &  0.18 $\pm$ 0.04 & -0.05 $\pm$ 0.10 (5) &  0.05 $\pm$ 0.09 (4) & -0.05 $\pm$ 0.09 (4) & -0.11 $\pm$ 0.05 (2) \\  
HD    26151  & 5383 &  0.33 $\pm$ 0.05 & -0.10 $\pm$ 0.06 (5) &  0.02 $\pm$ 0.05 (4) & -0.22 $\pm$ 0.11 (4) & -0.04 $\pm$ 0.02 (2) \\  
HD    26794  & 4920 &  0.07 $\pm$ 0.03 & -0.17 $\pm$ 0.10 (2) &  0.02 $\pm$ 0.05 (4) & -0.10 $\pm$ 0.02 (3) & -0.02 $\pm$  --- (1) \\  
HD    27894  & 4920 &  0.37 $\pm$ 0.03 & -0.17 $\pm$ 0.03 (2) & -0.00 $\pm$ 0.12 (4) & -0.16 $\pm$ 0.16 (2) &  0.01 $\pm$ 0.02 (2) \\  
HD    30295  & 5406 &  0.32 $\pm$ 0.04 & -0.15 $\pm$ 0.08 (5) &  0.01 $\pm$ 0.06 (4) & -0.16 $\pm$ 0.07 (3) &  0.02 $\pm$ 0.07 (2) \\  
HD    31452  & 5250 &  0.23 $\pm$ 0.04 & -0.18 $\pm$ 0.05 (5) & -0.01 $\pm$ 0.08 (4) & -0.11 $\pm$ 0.08 (2) & -0.01 $\pm$  --- (1) \\  
HD    31827  & 5608 &  0.48 $\pm$ 0.04 & -0.04 $\pm$ 0.06 (5) &  0.06 $\pm$ 0.08 (4) & -0.15 $\pm$ 0.15 (4) & -0.03 $\pm$ 0.03 (2) \\	
HD    35854  & 4901 & -0.04 $\pm$ 0.03 &  0.08 $\pm$ 0.16 (2) &  0.10 $\pm$ 0.04 (4) & -0.14 $\pm$ 0.08 (4) & \multicolumn{1}{c}{---} \\     
HD    37986  & 5503 &  0.30 $\pm$ 0.04 & -0.07 $\pm$ 0.06 (5) & -0.06 $\pm$ 0.08 (4) & -0.08 $\pm$ 0.08 (4) &  0.07 $\pm$ 0.07 (2) \\	
HD    39213  & 5473 &  0.45 $\pm$ 0.05 & -0.15 $\pm$ 0.08 (5) &  0.05 $\pm$ 0.08 (4) & -0.17 $\pm$ 0.06 (3) & -0.12 $\pm$ 0.05 (2) \\	
HD    39715  & 4741 & -0.10 $\pm$ 0.03 &  0.09 $\pm$ 0.18 (2) &  0.09 $\pm$ 0.07 (4) &  0.01 $\pm$ 0.08 (2) & -0.02 $\pm$ 0.00 (2) \\	
HD    43848  & 5161 &  0.43 $\pm$ 0.03 & -0.10 $\pm$ 0.07 (5) &  0.04 $\pm$ 0.10 (4) & -0.31 $\pm$ 0.11 (4) & -0.10 $\pm$ 0.03 (2) \\	
HD    77338  & 5346 &  0.41 $\pm$ 0.04 & -0.14 $\pm$ 0.11 (5) &  0.05 $\pm$ 0.12 (4) & -0.19 $\pm$ 0.12 (3) & -0.10 $\pm$ 0.00 (2) \\	
HD    81767  & 4966 &  0.22 $\pm$ 0.04 & -0.16 $\pm$ 0.09 (3) &  0.01 $\pm$ 0.10 (4) & -0.19 $\pm$ 0.06 (4) & -0.19 $\pm$ 0.05 (2) \\	
HD    82943  & 5929 &  0.23 $\pm$ 0.04 & -0.06 $\pm$ 0.06 (5) &  0.06 $\pm$ 0.10 (4) & -0.16 $\pm$ 0.12 (3) & -0.11 $\pm$ 0.14 (2) \\	
HD    86065  & 4938 &  0.09 $\pm$ 0.04 & -0.00 $\pm$ 0.18 (3) &  0.09 $\pm$ 0.06 (4) & -0.07 $\pm$ 0.03 (3) & -0.01 $\pm$ 0.16 (2) \\	
HD    86249  & 4957 &  0.12 $\pm$ 0.04 &  0.01 $\pm$ 0.16 (3) &  0.09 $\pm$ 0.06 (4) & -0.11 $\pm$ 0.08 (3) & -0.21 $\pm$ 0.02 (2) \\	
HD    87007  & 5282 &  0.29 $\pm$ 0.05 & -0.04 $\pm$ 0.13 (5) & -0.05 $\pm$ 0.05 (4) & -0.18 $\pm$ 0.13 (4) &  0.01 $\pm$ 0.08 (2) \\	
HD    90054  & 6047 &  0.29 $\pm$ 0.05 & -0.16 $\pm$ 0.09 (4) & -0.07 $\pm$ 0.10 (4) & -0.29 $\pm$ 0.04 (3) &  0.02 $\pm$ 0.05 (2) \\	
HD    91585  & 5144 &  0.25 $\pm$ 0.04 & -0.09 $\pm$ 0.06 (5) & -0.01 $\pm$ 0.07 (4) & -0.14 $\pm$ 0.03 (3) &  0.03 $\pm$ 0.11 (2) \\	
HD    91669  & 5278 &  0.44 $\pm$ 0.04 & -0.11 $\pm$ 0.09 (5) & -0.07 $\pm$ 0.13 (4) & -0.25 $\pm$ 0.11 (3) & -0.11 $\pm$ 0.19 (2) \\	
HD    93800  & 5181 &  0.49 $\pm$ 0.04 & -0.03 $\pm$ 0.08 (5) &  0.03 $\pm$ 0.10 (4) & -0.29 $\pm$ 0.15 (3) & -0.09 $\pm$ 0.03 (2) \\	
HD    94374  & 5000 & -0.10 $\pm$ 0.03 & -0.11 $\pm$  --- (1) &  0.12 $\pm$ 0.05 (4) &  0.28 $\pm$  --- (1) & \multicolumn{1}{c}{---} \\      
HD    95338  & 5175 &  0.21 $\pm$ 0.04 & -0.12 $\pm$ 0.06 (5) & -0.04 $\pm$ 0.07 (4) & -0.20 $\pm$ 0.09 (2) & -0.04 $\pm$ 0.01 (2) \\	
HD   104212  & 5833 &  0.13 $\pm$ 0.05 & -0.18 $\pm$ 0.08 (5) &  0.04 $\pm$ 0.06 (4) & -0.24 $\pm$ 0.08 (4) & -0.13 $\pm$ 0.04 (2) \\	
HD   107509  & 6102 &  0.03 $\pm$ 0.05 & -0.14 $\pm$ 0.05 (5) & -0.04 $\pm$ 0.06 (4) & -0.12 $\pm$ 0.05 (4) & -0.08 $\pm$ 0.12 (2) \\	
HD   120329  & 5617 &  0.31 $\pm$ 0.06 & -0.16 $\pm$ 0.05 (5) & -0.05 $\pm$ 0.07 (4) & -0.14 $\pm$ 0.10 (3) & -0.11 $\pm$ 0.09 (2) \\	
HD   143102  & 5547 &  0.16 $\pm$ 0.05 & -0.17 $\pm$ 0.05 (5) & -0.02 $\pm$ 0.05 (4) & -0.21 $\pm$ 0.05 (4) & -0.11 $\pm$ 0.06 (2) \\	
HD   148530  & 5392 &  0.03 $\pm$ 0.05 & -0.12 $\pm$ 0.06 (5) &  0.02 $\pm$ 0.06 (4) & -0.05 $\pm$ 0.15 (3) &  0.09 $\pm$ 0.19 (2) \\	
HD   149256  & 5406 &  0.34 $\pm$ 0.05 & -0.18 $\pm$ 0.05 (5) & -0.09 $\pm$ 0.06 (4) & -0.23 $\pm$ 0.06 (4) &  0.03 $\pm$ 0.02 (2) \\	
HD   149606  & 4976 &  0.20 $\pm$ 0.04 & -0.11 $\pm$  --- (1) &  0.05 $\pm$ 0.10 (4) & -0.17 $\pm$ 0.13 (3) & -0.16 $\pm$ 0.14 (2) \\	
HD   149933  & 5486 &  0.13 $\pm$ 0.06 & -0.06 $\pm$ 0.12 (4) & -0.06 $\pm$ 0.08 (4) &  0.23 $\pm$ 0.03 (2) &  0.29 $\pm$ 0.02 (2) \\	
HD   165920  & 5336 &  0.36 $\pm$ 0.04 & -0.10 $\pm$ 0.07 (5) &  0.01 $\pm$ 0.08 (4) & -0.16 $\pm$ 0.04 (4) & -0.11 $\pm$ 0.08 (2) \\	
HD   168714  & 5686 &  0.48 $\pm$ 0.05 & -0.18 $\pm$ 0.06 (5) & -0.00 $\pm$ 0.08 (4) & -0.18 $\pm$ 0.07 (3) & -0.13 $\pm$ 0.16 (2) \\	
HD   171999  & 5304 &  0.29 $\pm$ 0.04 & -0.18 $\pm$ 0.09 (5) & -0.02 $\pm$ 0.08 (4) & -0.13 $\pm$ 0.10 (4) & -0.09 $\pm$ 0.00 (2) \\	
HD   177374  & 5044 & -0.08 $\pm$ 0.03 &  0.03 $\pm$ 0.17 (2) & -0.05 $\pm$ 0.05 (4) & -0.03 $\pm$ 0.02 (2) & \multicolumn{1}{c}{---} \\      
HD   179764  & 5323 & -0.05 $\pm$ 0.04 & -0.11 $\pm$ 0.07 (5) & -0.08 $\pm$ 0.05 (4) & -0.08 $\pm$ 0.12 (4) &  0.11 $\pm$ 0.01 (2) \\	
HD   180865  & 5218 &  0.21 $\pm$ 0.04 & -0.15 $\pm$ 0.03 (4) &  0.06 $\pm$ 0.08 (4) & -0.12 $\pm$ 0.10 (4) &  0.02 $\pm$ 0.02 (2) \\	
HD   181234  & 5311 &  0.45 $\pm$ 0.04 & -0.09 $\pm$ 0.12 (5) &  0.08 $\pm$ 0.11 (4) & -0.18 $\pm$ 0.08 (3) & -0.04 $\pm$ 0.12 (2) \\	
HD   181433  & 4902 &  0.41 $\pm$ 0.04 &  0.02 $\pm$ 0.18 (3) & -0.09 $\pm$ 0.10 (4) & -0.17 $\pm$ 0.09 (2) & -0.08 $\pm$ 0.16 (2) \\	
HD   182572  & 5700 &  0.48 $\pm$ 0.03 & -0.15 $\pm$ 0.03 (5) &  0.03 $\pm$ 0.07 (4) & -0.19 $\pm$ 0.08 (4) & -0.11 $\pm$ 0.12 (2) \\	
HD   196397  & 5404 &  0.38 $\pm$ 0.05 & -0.16 $\pm$ 0.17 (5) &  0.03 $\pm$ 0.08 (4) & -0.19 $\pm$ 0.06 (4) & -0.04 $\pm$  --- (1) \\	
HD   196794  & 5094 &  0.06 $\pm$ 0.04 &  0.14 $\pm$ 0.05 (3) &  0.24 $\pm$ 0.05 (4) & -0.01 $\pm$ 0.12 (4) &  0.07 $\pm$  --- (1) \\  
HD   197921  & 4866 &  0.22 $\pm$ 0.04 & -0.22 $\pm$ 0.08 (3) &  0.00 $\pm$ 0.10 (4) & -0.16 $\pm$ 0.07 (4) & -0.03 $\pm$ 0.08 (2) \\	
HD   201237  & 4829 &  0.00 $\pm$ 0.04 & -0.22 $\pm$ 0.06 (2) & -0.05 $\pm$ 0.11 (4) & -0.27 $\pm$ 0.09 (3) & -0.10 $\pm$ 0.14 (2) \\	   
HD   209721  & 5503 &  0.28 $\pm$ 0.04 & -0.15 $\pm$ 0.13 (5) & -0.15 $\pm$ 0.10 (4) & -0.18 $\pm$ 0.11 (4) & -0.03 $\pm$ 0.01 (2) \\
HD   211706  & 6017 &  0.09 $\pm$ 0.07 & -0.10 $\pm$ 0.05 (4) &  0.00 $\pm$ 0.05 (4) & -0.16 $\pm$ 0.06 (2) &  0.04 $\pm$ 0.03 (2) \\
HD   213996  & 5314 &  0.33 $\pm$ 0.04 & -0.12 $\pm$ 0.08 (5) & -0.06 $\pm$ 0.05 (4) & -0.05 $\pm$ 0.12 (3) & -0.12 $\pm$ 0.10 (2) \\
HD   214463  & 5122 &  0.34 $\pm$ 0.04 & -0.10 $\pm$ 0.13 (4) & -0.09 $\pm$ 0.11 (4) & -0.21 $\pm$ 0.08 (4) &  0.05 $\pm$ 0.03 (2) \\
HD   218566  & 4849 &  0.28 $\pm$ 0.14 & -0.11 $\pm$ 0.07 (3) &  0.04 $\pm$ 0.12 (4) & -0.20 $\pm$ 0.13 (4) & -0.04 $\pm$ 0.07 (2) \\
HD   218750  & 5134 &  0.17 $\pm$ 0.04 & -0.18 $\pm$ 0.10 (4) &  0.05 $\pm$ 0.08 (4) & -0.21 $\pm$ 0.07 (2) &  0.11 $\pm$  --- (1) \\
HD   221313  & 5153 &  0.31 $\pm$ 0.05 & -0.15 $\pm$ 0.08 (5) &  0.02 $\pm$ 0.09 (4) & -0.14 $\pm$ 0.11 (3) & -0.10 $\pm$  --- (1) \\
HD   221974  & 5213 &  0.46 $\pm$ 0.04 & -0.01 $\pm$ 0.06 (4) &  0.03 $\pm$ 0.10 (4) & -0.12 $\pm$ 0.16 (3) & -0.17 $\pm$ 0.12 (2) \\
HD   224230  & 4873 & -0.08 $\pm$ 0.04 &  0.04 $\pm$ 0.27 (2) &  0.03 $\pm$ 0.07 (4) & -0.07 $\pm$ 0.02 (2) &  0.15 $\pm$ 0.09 (2) \\
HD   224383  & 5760 & -0.10 $\pm$ 0.05 & -0.06 $\pm$ 0.12 (5) & -0.04 $\pm$ 0.05 (4) & -0.08 $\pm$ 0.13 (4) &  0.13 $\pm$ 0.09 (2) \\

\label{Tab_final_abonds}

\end{longtable}

\twocolumn
\clearpage

\bibliographystyle{aa}


\end{document}